\pdfoutput=1
\documentclass[a4paper,11pt]{article}
\usepackage{amsmath,amssymb,amsfonts,graphicx}
\usepackage[noconfig]{refstyle}
\usepackage{cite}
\usepackage[dvipsnames]{xcolor}
\usepackage[hyperfootnotes=false, linktocpage=true, colorlinks, citecolor=blue, linkcolor=blue, urlcolor=Maroon, filecolor=Maroon]{hyperref}
\usepackage{geometry,longtable}
\usepackage[small,bf]{caption}
\usepackage{tikz}
\usetikzlibrary{shapes,arrows}
\tikzstyle{block} = [rectangle, draw, text width=7em, text centered, rounded corners, minimum height=3em]

\usepackage[normalem]{ulem}

\let\eqref=\relax
\newref{eq}{name={},Name={Eq.~},names={eqs.~},Names={Eqs.~},rngtxt={-},refcmd=(\ref{#1})}
\newref{tab}{name={},Name={Table~},names={tables~},Names={Tables~}}
\newref{sec}{name={},Name={Section~},names={sections~},Names={Sections~}}
\newref{fig}{name={figure~},Name={Figure~},names={figures~},Names={Figures~}}
\numberwithin{equation}{section}

\newcommand{\be}{\begin{equation}}
\newcommand{\ee}{\end{equation}}

\newcommand{\dps}[1]{\mathrm{dP}_{#1}}
\def\fnote#1#2{\begingroup\def\thefootnote{#1}\footnote{#2}
     \addtocounter{footnote}{-1}\endgroup}

\begin{document}

\vspace{1cm}

\title{\vspace{40pt}
       {\Large \bf Machine Learning Line Bundle Cohomology}\\[2mm]
}

\author{
Callum R. Brodie${}^{1}$,
Andrei Constantin${}^{2}$,
Rehan Deen${}^{1}$,
Andre Lukas${}^1$
}
\date{}
\maketitle
\begin{center} {\small \vskip -3mm
  ${}^1${\it Rudolf Peierls Centre for Theoretical Physics, University of Oxford,\\ Parks Road, Oxford OX1 3PU, UK}\\[4mm]
  ${}^2${\it Pembroke College, University of Oxford, OX1 1DW, UK\\
Mansfield College, University of Oxford, OX1 3TF, UK}}\\

  \fnote{}{callum.brodie@physics.ox.ac.uk}
 \fnote{}{andrei.constantin@mansfield.ox.ac.uk}
 \fnote{}{rehan.deen@physics.ox.ac.uk}
 \fnote{}{lukas@physics.ox.ac.uk} 
\end{center}

\begin{abstract}
\noindent We investigate different approaches to machine learning of line bundle cohomology on complex surfaces as well as on Calabi-Yau three-folds. Standard function learning  based on simple fully connected networks with logistic sigmoids is reviewed and its main features and shortcomings are discussed. It has been observed recently that line bundle cohomology can be described by dividing the Picard lattice into certain regions in each of which the cohomology dimension is described by a polynomial formula. Based on this structure, we set up a network capable of identifying the regions and their associated polynomials, thereby effectively generating a conjecture for the correct cohomology formula. For complex surfaces, we also set up a network which learns certain rigid divisors which appear in a recently discovered master formula for cohomology dimensions.
\end{abstract}

\thispagestyle{empty}
\setcounter{page}{0}
\newpage

\tableofcontents
\newpage 

\section{Introduction}\seclabel{intro}
Techniques of machine learning have recently been introduced into string theory~\cite{He:2017set,He:2017aed,Ruehle:2017mzq} and have been applied to a range of problems~\cite{Bull:2018uow,Erbin:2018csv,He:2018jtw,Cole:2018emh,Halverson:2019tkf}. Broadly, these applications can be divided into two classes, firstly, those which are attempting to facilitate difficult mathematical calculations required within string theory (``replacing the Mathematician") and, secondly, those which employ machine learning techniques to deal with the vast amount of data string theory provides (``replacing the string theorist"). In the present paper, we are interested in problems within the first class, specifically in applications of machine learning to the problem of computing (line) bundle cohomology dimensions on complex manifolds. 

The computation of (line) bundle cohomology is required in various types of string compactifications, notably in heterotic compactifications, in type II compactifications with brane flux and in F-theory, and is, hence, ubiquitous in string theory. General mathematical methods to compute bundle cohomology exist, usually based on \v{C}ech cohomology, sometimes combined with spectral sequence techniques, but these methods are algorithmic and often computationally intense. The complicated nature of these calculations frequently stands in the way of attempting a ``bottom-up" model building approach in constructions where bundles are involved. A better understanding of bundle cohomology, while being a worthwhile mathematical goal in its own right, is, therefore, also crucial for more systematic string model building.

In this paper, we will be concerned with line bundle cohomology dimensions on complex manifolds, specifically complex surfaces which may arise as GUT surfaces within F-theory, and Calabi-Yau three-folds which underly heterotic compactifications. Our main goal is to use machine learning in order to gain a better insight into the structure of line bundle cohomology on such manifolds. Throughout we will rely on supervised machine learning only.

After a brief introduction of the mathematical background in Section~\ref{background}, we review the standard approach to machine learning line bundle cohomology in Section~\ref{fully}. We observe that this approach comes with a number of disadvantages. For one, even relatively successful networks which reproduce 90\% (say) of the required cohomology dimensions correctly are not of much practical use in string model building, where a single construction might involve $20$ or $30$ line bundles, each of which needs to be known exactly in order to obtain correct predictions. Also, a ``black-box" network, merely capable of predicting individual dimensions, does not provide any information about the underlying mathematical structure.

In Section~\ref{learnform} we will, therefore, discuss a different method based on recent insight into line bundle cohomology. The first observation that line bundle cohomology on Calabi-Yau three-folds can be captured in closed-form expressions appeared in the study of the tetra-quadric manifold in Refs.~\cite{Constantin:2018otr, Buchbinder:2013dna}.
In Ref.~\cite{Constantin:2018hvl}, it was found empirically for a number of Calabi-Yau three-fold examples, that the Picard group (which labels line bundles) splits into different regions, frequently cones, in each of which the cohomology dimensions are given by a cubic polynomial. Recently, in Ref.~\cite{Larfors:2019sie}, this behaviour was investigated and confirmed for a larger class of Calabi-Yau three-folds. Ref.~\cite{Klaewer:2018sfl} makes a similar observation for other examples, including complex surfaces (where the relevant polynomials are quadratic), relying on machine-learning techniques. Our two companion papers~\cite{papermath,paperex} provide a proof for the existence of such formulae which applies to the zeroth cohomology of certain classes of complex surfaces, including del Pezzo surfaces, Hirzebruch surfaces and compact toric surfaces. These papers also present a ``master formula" for zeroth cohomology on complex surfaces which involves certain rigid divisors of the surface.

Our approach in Section~\ref{learnform} is based on the observation that knowledge of the regions in the Picard lattice is sufficient to extract the complete information. Indeed, given such a region it is easy to fit a polynomial of the relevant degree (quadratic for complex surfaces, cubic for three-folds) to a collection of points in the interior of the region. We will, therefore, set up a network which can be trained to identify those regions. Combined with a simple fit algorithm and a few other straightforward steps this network works essentially as a conjecture generator, predicting the relevant cohomology formula for the space under consideration. Our main goal is a proof of concept,  demonstrating that this approach to conjecture-generating does indeed work. 

In Section~\ref{learnmaster} we attempt a different approach based on the master formula for the zeroth cohomology of line bundles on surfaces found in Refs.~\cite{papermath,paperex}. We will set up a network capable, at least in some cases, of learning the rigid divisors which enter the master formula.  As before, our main goal is a proof of concept. We conclude in Section~\ref{conclusion}.

\section{Background}\label{background}
In this section, we collect some of the mathematical facts underlying our subsequent machine learning applications and set up our notation. Details, including proofs, can be found in Refs.~\cite{Constantin:2018hvl,papermath,paperex}, and general mathematical background can, for example, be found in Refs.~\cite{hartshorne,gh,huebsch}.

\subsection{Mathematical setting}
In the following, we consider complex manifolds $X$ of (complex) dimension $d$, with our examples taken from complex surfaces ($d=2$) and three-folds ($d=3$).
The set of line bundles (up to isomorphisms) on $X$ is called the {\it Picard group}, ${\rm Pic}(X)\cong\mathbb{Z}^h$, whose rank, $h$, equals the Hodge number $h^{1,1}(X)$ for all examples under consideration. We will make use of the divisor line bundle correspondence and write a line bundle $L\rightarrow X$ which corresponds to a divisor $D\subset X$ as $L={\cal O}_X(D)$. More practically, we usually choose a basis ${\cal D}_i$ of (classes of) divisors, where $i=1,\ldots ,h$ and for $D=k_1{\cal D}_1+\cdots  k_h{\cal D}_h$ we write the associated line bundle as $L={\cal O}_X({\bf k})$ where ${\bf k}=(k_1,\ldots ,k_h)\in\mathbb{Z}^h$.\\[2mm]
Our main interest is in the cohomology dimensions $h^q(L)={\rm dim}\, H^q(X,L)\in\mathbb{Z}^{\geq 0}$, where $q=0,\ldots ,d$, of the line bundles on $X$. By means of machine learning, we are trying to understand the dependence of these cohomology dimensions, for a given manifold $X$,  on the choice of line bundle, that is, on the integer vector ${\bf k}$. This means our training data is of the form
\begin{equation}
 \mathbb{Z}^h\ni{\bf k}\;\longrightarrow\; h^q({\cal O}_X({\bf k}))\in\mathbb{Z}^{\geq 0}\; .
\end{equation} 
There are two more general facts about bundle cohomology which are relevant. One is Serre duality which implies that
\begin{equation}
 h^q(L)=h^{d-q}(K_X\otimes L^*)\; ,
\end{equation}
where $K_X$ is the canonical line bundle of $X$ and $L^*={\cal O}_X(-{\bf k})$ is the dual bundle. The other fact is the index theorem
\begin{equation}
 {\rm ind}(L)=\sum_{q=0}^d(-1)^qh^q(L)=\int_X{\rm Td}(X)\wedge{\rm ch}(L)\; , \eqlabel{ASI}
\end{equation}
where ${\rm Td}(X)$ is the Todd class of the manifold $X$ and ${\rm ch}(L)$ is the Chern character of $L={\cal O}_X({\bf k})$.  Both quantities can be computed explicitly for a given manifold $X$ and line bundle $L\rightarrow X$ and, hence, the right-hand-side of Eq.~\eqref{ASI} is known explicitly. It is, in fact, given by a polynomial of degree $d$ in the line bundle integers $k_i$. 

For complex surfaces we have three cohomology dimensions $h^q(L)$, where $q=0,1,2$, and  know\-ledge of $h^0(L)$ for all line bundles $L\rightarrow X$ determines the other cohomology dimensions, via Serre duality and the index theorem,. For three-folds we have four dimensions $h^q(X)$, where $q=0,1,2,3$, and we require knowledge of two cohomology dimensions, say $h^0(L)$ and $h^1(L)$, for all line bundles $L\rightarrow X$, with the other cohomologies then determined by Serre duality.\\[2mm]
It has been found empirically~\cite{Constantin:2018hvl,Klaewer:2018sfl,paperex} for a number of example spaces, including surfaces and three-folds, by explicit computation of line bundle cohomology dimensions for many line bundles ${\cal O}_X({\bf k})$, that there exist piecewise polynomial formulae for $h^q({\cal O}({\bf k}))$.  More specifically, the Picard group ${\rm Pic}(X)\cong \mathbb{Z}^h=\bigcup_\alpha R_\alpha$ splits into disjoint regions $R_\alpha$, such that $h^q({\cal O}({\bf k}))=p_\alpha({\bf k})$ for all ${\bf k}\in R_\alpha$, where $p_\alpha$ is a polynomial of degree $d$ in the components $k_i$ of ${\bf k}$. In Section~\ref{learnform} we will see how knowledge of this basic structure facilitates machine learning of line bundle cohomology and leads to a conjecture-generating network. 

There is one more empirical observation about the structure of these formulae which is worth mentioning in the present context. If the manifold $X$ has an ample anti-canonical bundle $-K_X$ then the regions $R_\alpha$ have dimension $h$ and the cohomology dimensions match ``continuously" at the region boundaries. On other hand, if $-K_X$ is not ample (in particular, if it is trivial and, hence, the space is a Calabi-Yau manifold) there are regions $R_\alpha$ of dimension $h$ as well as of lower dimension and cohomology dimensions can jump discontinuously at region boundaries. As we will see, these features have implications for machine learning of cohomology dimensions. \\[2mm]
For the case of the zeroth cohomology on surfaces, the origin of piecewise quadratic formulae has been explained - and in many cases proven - in Refs.~\cite{papermath,paperex}. The key is a map $D\rightarrow \tilde{D}$ between divisors defined by
\begin{equation}
 \tilde{D}=D-\sum_{C\in{\cal I}}\;\theta(-C\cdot D)\,{\rm ceil}\left(\frac{C\cdot D}{C^2}\right) C\; , \eqlabel{master}
\end{equation} 
 where $\theta$ and  ${\rm ceil}$ are the Heaviside and ceiling functions, respectively, the dot denotes the intersection form on $X$ and the sum runs over the set ${\cal I}$ of (irreducible) divisors with negative self-intersection. It has been shown that for any divisor $D$ in the effective cone (that is, for any $D$ with $h^0(D)>0$) the zeroth cohomology is unchanged under the above map, that is, $h^0(\tilde{D})=h^0(D)$. In addition, it can be shown~\cite{papermath,paperex}, that for many surfaces, which include del Pezzo surfaces and Hirzebruch surfaces, there is a vanishing theorem (Kodaira vanishing or one of its variants) which applies to $\tilde{D}$ and which states that $h^q({\cal O}(\tilde{D}))=0$ for $q>0$. This means from Eq.~\eqref{ASI} that $h^0({\cal O}(\tilde{D}))$ can be computed from the index theorem and it follows that
 \begin{equation}
  h^0({\cal O}(D))={\rm ind}(\tilde{D})\; . \eqlabel{h0ind}
 \end{equation}
 Since the index is always a quadric in the line bundle integers $k_i$ this explains the structure  of the cohomology formulae for $h^0$ and provides a practical way of deriving them by identifying the divisors ${\cal I}$ summed over in Eq.~\eqref{master}.
 
For some surfaces, including some toric surfaces, it can be necessary to apply the formula~\eqref{master} multiple times in succession to arrive at a new divisor for which a suitable vanishing theorem applies. In such cases, the zeroth cohomology can still be expressed as the index of another divisor, but its relation to the original divisor becomes more complicated and amounts to iterating Eq.~\eqref{master}.
  
In the following, we will refer to Eq.~\eqref{master} as the ``master formula" for cohomology and Section~\ref{learnmaster}  discusses how this formula can be used in the context of machine learning.\\[2mm]
Unfortunately, at present, there are no analogous master formulae known for higher cohomologies or for three-folds. Nevertheless, the study of examples suggests the existence of piecewise polynomial formulae in those cases as well and we will rely on and confirm  this empirical fact in some of our machine learning applications. Let us now discuss the main classes of example manifolds $X$ which we will use throughout the paper.

\subsection{Del Pezzo surfaces}
Del Pezzo surfaces $\dps{r}$, where $r=0,\ldots ,8$, are defined as the complex projective plane $\mathbb{P}^2$ blown up in $r$ (generic) points. (In particular, $\dps{0}=\mathbb{P}^2$ is the projective plane and, hence, somewhat trivial.) The rank of the Picard lattice is $h=h^{1,1}(\dps{r})=r+1$ and 
a basis of divisor classes is given by $({\cal D}_i)=(l,e_1,\ldots ,e_r)$, where $l$ is the hyperplane class of $\mathbb{P}^2$ and $e_i$ are the exceptional divisor classes associated to the blow-ups. The intersection form is determined by the relations
\begin{equation}
 l\cdot l=1\;,\qquad l\cdot e_i=0\;,\qquad e_i\cdot e_j=-\delta_{ij}\; . \eqlabel{intdP}
\end{equation} 
Line bundles are labelled by an integer vector ${\bf k}=(k_0,k_1,\ldots ,k_r)\in\mathbb{Z}^{r+1}$ and are written as ${\cal O}_{\dps{r}}({\bf k})={\cal O}_{\dps{r}}(k_0l+k_1e_1+\cdots +k_re_r)$. The canonical bundle and Serre duality take the form
\begin{equation}
 K_{\dps{r}}=-3l+\sum_{i=1}^r e_i\; ,\qquad h^2({\cal O}_{\dps{r}}({\bf k}))=h^0({\cal O}_{\dps{r}}(-k_0-3,-k_1+1,\ldots ,-k_r+1))
\end{equation}
and, hence, del Pezzo surfaces have an ample anti-canonical bundle. The index theorem reads explicitly 
\begin{equation}
{\rm ind}({\cal O}_{\dps{r}}({\bf k}))=h^0({\cal O}_{\dps{r}}({\bf k}))-h^1({\cal O}_{\dps{r}}({\bf k})) +h^2({\cal O}_{\dps{r}}({\bf k}))=1+\frac{1}{2}k_0(k_0+3)+\frac{1}{2}\sum_{i=1}^r k_i(1-k_i)\; .
\eqlabel{inddP}
\end{equation}
As discussed above, these two results can be used to determine $h^1$ and $h^2$ once $h^0$ is known for all line bundles. 

Line bundle cohomology dimensions on del Pezzo surfaces can be calculated by three algorithmic methods:
\begin{itemize}
 \item The del Pezzo surfaces $\dps{r}$ for $r=0,1,2,3$ have a toric realisation. For those cases the algorithm of Ref.~\cite{Blumenhagen:2010pv} which computes line bundle cohomology on toric spaces can be used.
 \item All del Pezzo surfaces $\dps{r}$ have realisations as (favourable) complete intersections in products of projective spaces~\cite{huebsch} so that the algorithm of Refs.~\cite{Anderson:2007nc,Gray:2007yq,Anderson:2008uw,He:2009wi,Anderson:2009mh} can be used.
 \item Ref.~\cite{Blumenhagen:2008zz} provides an algorithm to compute line bundle cohomology dimensions on all del Pezzo surfaces which is based on counting certain polynomials on $\mathbb{P}^2$.
\end{itemize} 
We have used all three methods in order to obtain the required training data. We note that for all cases we have checked the three methods agree wherever they overlap. As explained above, this data can be used to extract analytic, piecewise quadratic formulae for cohomology dimensions by ``eyeballing", although this process can be tedious.

As a simple example, the so-obtained formula for the zeroth cohomology of line bundles on $\dps{1}$ is given by
\begin{equation}
 h^0({\cal O}_{\dps{1}}(k_0,k_1))=\left\{\begin{array}{lll}\displaystyle\frac{1}{2}(k_0+1)(k_0+2)&&k_0\geq 0\,,\;k_1\geq 0\\[8pt]
                          \displaystyle\frac{1}{2}(k_0+1)(k_0+2)+\frac{1}{2}k_1(1-k_1)&&k_1<0\,,\;k_0>0\,,\;k_0+k_1\geq 0\\[12pt]
                          \,0&&\mbox{otherwise}\end{array}\right. \eqlabel{dP1h0}
\end{equation}  
As is evident there are three regions $R_\alpha$ is this case. In line with our general discussion, these regions are two-dimensional and cohomology dimensions match continuously at the boundaries. The second row corresponds to the (nef) cone where $h^0$ is given by the index, as comparison with Eq.~\eqref{inddP} confirms. The formula in the first row can be obtained by combining the master formula~\eqref{master} with Eq.~\eqref{h0ind} and inserting for $C$ the single divisor with self-intersection $-1$, namely $e_1$. A more detailed discussion of these issues and further examples can be found in Ref.~\cite{paperex}.  Our goal in Section~\ref{learnform} will be to conjecture formulae such as Eq.~\eqref{dP1h0} from machine learning.\\[2mm]

\subsection{Complete intersection manifolds}
Our three-fold examples will be taken from the class of complete intersection (CI) manifolds in products of projective spaces~\cite{Candelas:1987kf,Green:1986ck,huebsch}. Underlying the construction is an ambient space ${\cal A}=\mathbb{P}^{n_1}\times\cdots\times\mathbb{P}^{n_m}$, a product of complex projective spaces. The manifold $X\subset {\cal A}$ is defined as the common zero locus of homogeneous polynomials $P_1,\ldots ,P_K$. Their degrees of homogeneity are encoded in a configuration matrix
\begin{equation}
 X\in\left[\begin{array}{c|ccc}\mathbb{P}^{n_1}&q^1_1&\cdots&q^1_K\\\vdots&\vdots&&\vdots\\\mathbb{P}^{n_m}&q^m_1&\cdots&q^m_K\end{array}\right]
 \qquad\begin{array}{c}{\cal D}_1\\\vdots\\{\cal D}_m\end{array}
\end{equation} 
Specifically, the entry $q_a^i$ of this matrix is the degree of the polynomial $P_a$ in the homogeneous coordinates of the $i^{\rm th}$ projective space. The complex dimension of the space is given by $d=\sum_{i=1}^mn_i-K$ and we are interested in the case of CI surfaces ($d=2$) and CI three-folds ($d=3$). The ${\cal D}_i$ listed after the configuration matrix are the divisor classes dual to the standard K\"ahler forms of the projective space (restricted to $X$) and we will focus on favourable cases, where these ${\cal D}_i$ span the entire fourth homology of $X$. For such cases, the rank of the Picard group is $h=h^{1,1}(X)=m$ and line bundles are labelled by integer vectors ${\bf k}=(k_1,\ldots ,k_m)\in\mathbb{Z}^m$ and denoted by ${\cal O}_X({\bf k})={\cal O}_X(k_1{\cal D}_1+\cdots +k_m{\cal D}_m)$. 

The anti-canonical bundle of such a complete intersection is given by
\begin{equation}
 -K_X=\sum_{i=1}^m(n_i+1-\sum_{a=1}^Kq_a^i){\cal D}_i\;.
\end{equation} 
This means, by choosing polynomial degrees, we can create CI manifolds with ample anti-canonical bundle (such as del Pezzo surfaces), CI Calabi-Yau manifolds (CICYs) if we choose $\sum_{a=1}^Kq^i_a=n_i+1$ for all $i=1,\ldots ,m$, so that $K_X=0$, or CI manifolds with an ample canonical bundle.

Line bundle cohomology on CI manifolds can be computed using the algorithm of Refs.~\cite{Anderson:2007nc,Gray:2007yq,Anderson:2008uw,He:2009wi,Anderson:2009mh} which relies on the Bott-Borel-Weil representation of cohomology on projective spaces combined with spectral sequence methods.  We expect line bundle cohomology dimensions on CI manifolds to be described by piecewise polynomial formulae, with quadratic polynomials for CI surfaces and cubic polynomials for CI three-folds. For the case of three-folds, this has first been shown in Ref.~\cite{Constantin:2018hvl} where several examples have been given.

As an illustration, consider the bi-cubic CICY three-fold in ${\cal A}=\mathbb{P}^2\times\mathbb{P}^2$, defined by the configuration matrix
\begin{equation}
 X\in\left[\begin{array}{c|c}\mathbb{P}^2&3\\\mathbb{P}^2&3\end{array}\right]\; . \eqlabel{confbicubic}
\end{equation} 
Line bundles ${\cal O}_X({\bf k})$ are labelled by a two-dimensional integer vector ${\bf k}=(k_1,k_2)$. Since the cohomology dimensions are invariant under the exchange $k_1\leftrightarrow k_2$ we can assume that $k_1\leq k_2$ without loss of generality. Under this assumption, the analytic formulae for the zeroth and first cohomology are~\cite{Constantin:2018hvl}
\begin{align}
\eqlabel{H0:bicubic}
h^0({\cal O}_X({\bf k}))  &  = 
	\begin{cases}
	\displaystyle \frac{1}{2}(1+k_2)(2+k_2)~, & k_1=0,~k_2\geq 0\\[8pt] 
	{\rm ind}({\cal O}_X({\bf k}))~, & k_1,k_2>0\\[8pt]
	0 & \text{otherwise}
	\end{cases}
\\[8pt]
\eqlabel{H1:bicubic}
h^1({\cal O}_X({\bf k}))  & = 
	\begin{cases}
	\displaystyle \frac{1}{2} (-1+ k_2) (-2 + k_2) ~, & k_1=0,~k_2>0 \\[8pt] 
	\displaystyle-{\rm ind}({\cal O}_X({\bf k}))~, & k_1<0,~k_2>-k_1\\[8pt]
	0 & \text{otherwise}~,
	\end{cases}
\end{align}
with the index given by ${\rm ind}({\cal O}_X({\bf k}))=\frac{3}{2} (k_1 + k_2) (2 + k_1 k_2)$. In line with our general discussion, we have two-dimensional regions as well as one-dimensional ones, the latter along the positive coordinate axis. Moreover, it is evident from the above formulae that cohomology dimensions jump discontinuously across boundaries. As for surfaces, we would like to be able to conjecture formulae such as the above from machine learning. 

\subsection{Set-up for machine learning}
In summary, we are interested in exploring machine learning of line bundle cohomology on manifolds $X$ with complex dimensions $d=2,3$, of the type introduced above. Line bundles on these manifolds are labelled by integer vectors ${\bf k}\in\mathbb{Z}^h$ with components $k_i$ and are denoted ${\cal O}_X({\bf k})$. The cohomology dimensions $h^q({\cal O}_X({\bf k}))\in\mathbb{Z}^{\geq 0}$, where $q=0,\ldots ,d$, can be explicitly computed using the various algorithmic methods outlined above. This leads to our training/validation data which is of the form
\begin{equation}
 \mathbb{Z}^h\ni{\bf k}\,\longrightarrow\;h^q({\cal O}_X({\bf k}))\in\mathbb{Z}^{\geq 0}. \eqlabel{traindata}
\end{equation} 
In practice, this data can only be obtained algorithmically for relative small values of $|k_i|$. It will be taken from a ``training box" defined by $|k_i|\leq k_{\rm max}$, where $k_{\rm max}$ varies from $5$ to $20$, depending on the manifold. 

As discussed, there is evidence - and proofs in some cases - that cohomology dimensions on these spaces are described by formulae which are piecewise polynomial, with polynomials of degree $d$. For this reason, it will sometimes be useful to modify the training data to
\begin{equation}
\begin{array}{rcccl}
 (k_i,k_ik_j)_{i\leq j}&\longrightarrow& h^q({\cal O}_X({\bf k}))&\mbox{for}&d=2\\
  (k_i,k_ik_j,k_ik_jk_l)_{i\leq j\leq l}&\longrightarrow& h^q({\cal O}_X({\bf k}))&\mbox{for}&d=3\; .
\end{array}  \eqlabel{tdlin}
\end{equation}
By providing all monomials up to degree $d$ in $k_i$ as an input the problem is effectively converted into a piecewise linear one. 

A common measure for how successfully a trained network performs is the mean square loss on the validation set. In the case of function approximation, a mean square loss translates into a typical accuracy with which the function in question is approximated. For our application to bundle cohomology it makes sense to introduce a different and often more stringent measure of success. In practice, we are not necessarily satisfied with cohomology dimensions approximated by the network within, say, a few percent. We would like the network to predict the exact cohomology dimensions, after rounding to the nearest integer. We will, therefore, measure the success of training by the percentage of cohomology dimensions within the training box which are correctly reproduced after rounding. 

Our networks have been realised with the Mathematica machine learning suite, and the fully-connected networks in the next section as well as the networks that learn cohomology formulae have also been realised with TensorFlow/Keras. Unless stated otherwise, training will be accomplished using the ADAM optimiser~\cite{Kingma:2014vow}.

In the next section we will consider standard function-learning with fully connected networks and their application to line bundle cohomology. This section is mainly intended as a warm-up and as an opportunity to draw attention to the various disadvantages of this approach. The subsequent sections will be devoted to a more tailored approach to machine learning of line bundle cohomology, which is based on the emerging mathematical structure.

\section{Simple fully connected networks}\label{fully}
In this section, we attempt function-learning of line bundle cohomology with simple one or two hidden layer, fully-connected networks. 

\subsection{Network structure}
\begin{figure}[h!!!]
\begin{center}
\includegraphics[width=13cm]{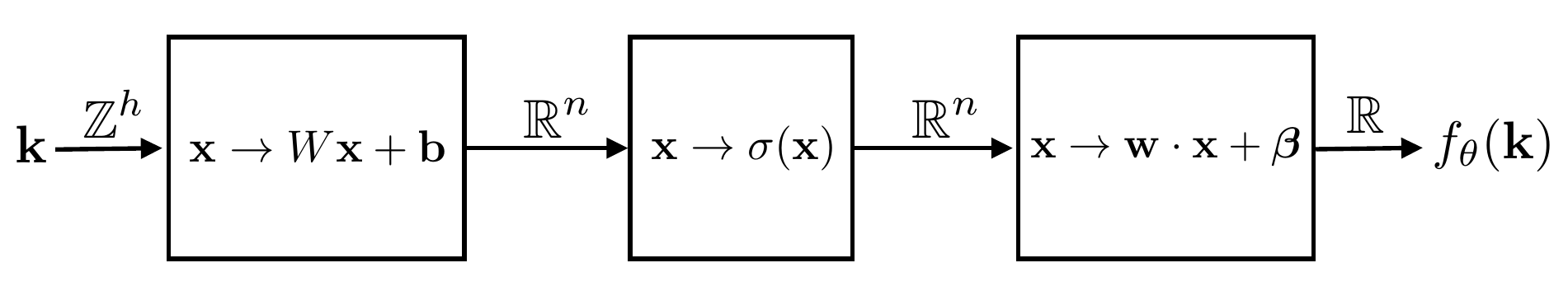}
\caption{\sf A simple one hidden layer fully connected network for function learning of line bundle cohomology. The hidden linear layer consists of $n$ neurons, the output layer of a single neuron. Here, $\sigma(x)=(1+\exp(-x))^{-1}$ is the logistic sigmoid function and $\theta=(W,{\bf b},{\bf w},{\boldsymbol\beta})$ collectively denotes all the weights and biases of the network.}
\label{fig:l1}
\end{center}
\vskip -5mm
\end{figure}

The structure of the one hidden layer network is shown in Fig.~\ref{fig:l1}.

The function $f_\theta$ represented by this network is explicitly given by
\begin{equation}
 f_\theta({\bf k})=\sum_{i=1}^n\sum_{j=1}^h\left(w_i\sigma(W_{ij}k_j+b_i)+\beta_i\right)\;,\qquad \sigma(x)=\frac{1}{1+\exp(-x)}\;,
\end{equation} 
and $\theta=(W,{\bf b},{\bf w},{\boldsymbol\beta})$ collectively denotes all the weights and biases of the network. The number, $n$, of neurons in the hidden layer will be chosen for each example, in view of optimising the fit. The universal approximation theorem~\cite{uap} asserts that the above series, for appropriate choices of weights and biases, can be made to converge uniformly to any continuous function on a hypercube in $\mathbb{R}^h$, as $n\rightarrow\infty$.

The functions under consideration are of course only defined on the discrete set $\mathbb{Z}^h$ and there is always a continuous function on $\mathbb{R}^h$ which matches all discrete values. Hence, it is, in principle, possible to approximate line bundle cohomology on any space and within the training box to arbitrary accuracy. In practice, this is of course limited and the continuity properties of the discrete function can impact on the quality of the approximation. As we have already mentioned, line bundle cohomology on manifolds with ample anti-canonical bundle is ``continuous" in the sense that the polynomials describing cohomology in each region match at the boundaries. For other cases, including for Calabi-Yau manifolds, on the other hand, cohomology can jump across the boundaries of regions. We can expect that these discontinuities will decrease the quality of the approximation. 

\begin{figure}[h!!!]
\begin{center}
\includegraphics[width=13cm]{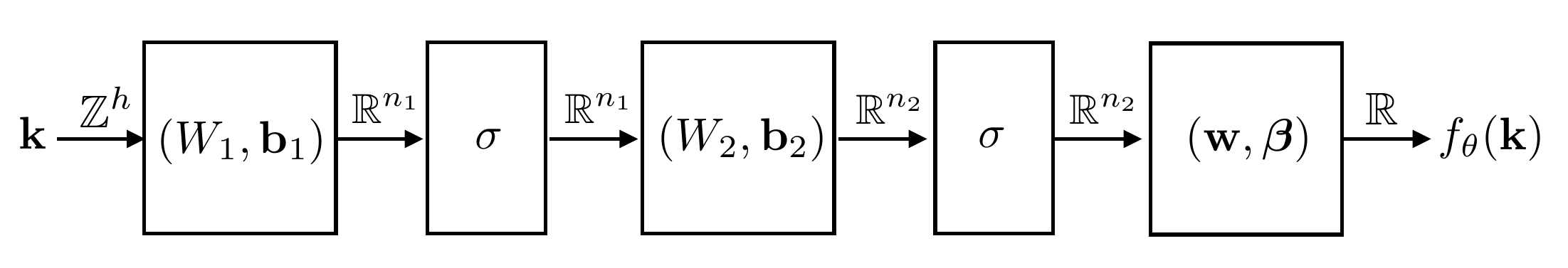}
\caption{\sf A two hidden layer fully connected network for function learning of line bundle cohomology. The two hidden linear layers consist of $n_1$ and $n_2$ neurons, respectively. Here, $\sigma(x)=(1+\exp(-x))^{-1}$ is the logistic sigmoid function and $\theta=(W_1,{\bf b}_1,W_2,{\bf b}_2,{\bf w},{\boldsymbol\beta})$ collectively denotes all the weights and biases of the network.}
\label{fig:l2}
\end{center}
\vskip -5mm
\end{figure}

For completeness, we will also consider the approximation of line bundle cohomology by fully connected networks with two hidden layers, of the type shown in Fig.~\ref{fig:l2}.

\subsection{Surface examples}
\begin{figure}[h!!!]
\begin{center}
\includegraphics[width=15cm]{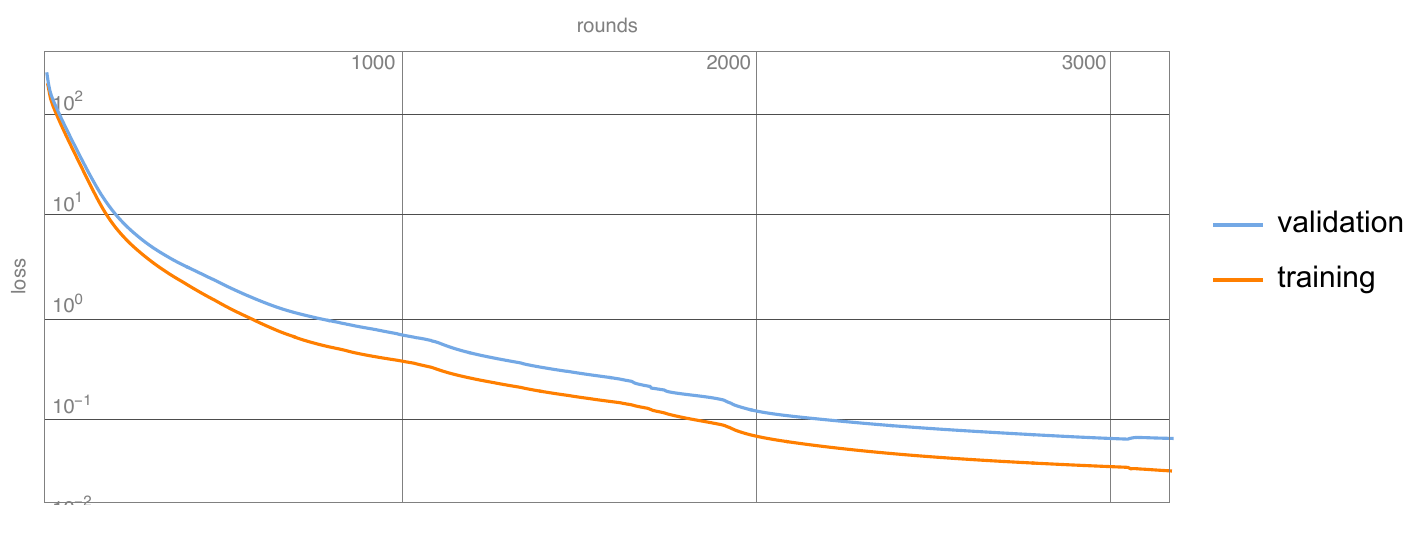}
\caption{\sf Loss as a function of training rounds for the zeroth cohomology on $\dps{2}$ and a single hidden layer network with $n=64$ neurons (Mathematica).}
\label{fig:dP2train}
\end{center}
\vskip -5mm
\end{figure}

For our surfaces examples, we consider the zeroth cohomology on the first three del Pezzo surfaces $\dps{r}$, for $r=1,2,3$, and we begin with a single hidden layer network as in Fig.~\ref{fig:l1}. We have used $n=64$ neurons in the hidden layer and, for $r=1,2,3$, respectively, we have selected training boxes of size $k_{\rm max}=15,10,10$ with a fraction of $0.5,\,0.2,\,0.1$ of cohomologies within the training box randomly selected for training, using 70\% as a training set and 30\% as a validation set. Training times are below a minute on a single CPU. A typical training curve, for the case $\dps{2}$ is shown in Fig.~\ref{fig:dP2train}.
In all three cases, the final validation loss is below $0.1$. 

The success rate, that is the fraction of cohomologies within the training box correctly reproduced after rounding to the nearest integer, is $0.97,\,0.94,\,0.93$ for $r=1,2,3$, respectively. 

In conclusion, machine learning of line bundle cohomologies on surfaces can be successfully accomplished with simple one hidden layer networks with $64$ neurons, achieving success rates larger than $0.9$.\\[2mm] 
It is well-known that neural networks do not typically provide reliable predictions outside the training box and this is true in the present case. For example, if we use the network for $\dps{2}$, with training box $k_{\rm max}=10$, to predict cohomologies outside the training box and with $|k_i|\leq 15$ the success rate decreases to $0.7$, down from $0.94$ within the training box.\\[2mm]
Similar results can be obtained from networks with two hidden layers, as in Fig.~\ref{fig:l2}. We use the same training boxes and training/validation sets as for the one layer networks, as well as network widths $(n_1,n_2)=(18,8),\,(16,8),\,(32,16)$ for $r=1,2,3$. This leads to success rates of $0.98,\,0.94,\,0.98$ for $r=1,2,3$, respectively.

\subsection{Three-fold examples}
We will consider two three-folds examples, the bi-cubic and the tetra-quadric CICY, but similar results apply to other CICYs. It is, of course, possible to fit to more than one or all four cohomology dimensions simultaneously, but, for simplicity, we focus on fitting one of them, typically $h^0$ or $h^1$, at a time.

As we will see, cohomology dimensions for three-folds are more difficult to fit than those for surfaces, essentially because cohomology dimensions increase cubicly in ${\bf k}$ and grow to much larger values. Reproducing these large dimensions exactly (after rounding) is a challenge, at least for the relatively simple networks considered here. In addition to the steep increase of cohomology dimensions in some regions, the cohomology is identically zero in other regions which often take up a significant part of the training box. Dealing with this is simplified by adding a ramp layer, which acts as
\begin{equation}
 x\rightarrow \left\{\begin{array}{lll}0&\mbox{for}&x<0\\x&\mbox{for}&x\geq 0\end{array}\right.\; .
\end{equation} 
This layer simply cuts off any negative values which we know cannot arise. Throughout this subsection, we will consider the one and two layer networks in Figs.~\ref{fig:l1} and \ref{fig:l2} with a ramp added on to the end. \\[2mm]
Our first example is the bi-cubic CICY mentioned earlier and defined by the configuration matrix
\begin{equation}
 X\in\left[\begin{array}{c|c}\mathbb{P}^2&3\\\mathbb{P}^2&3\end{array}\right]\; .
\end{equation} 
We begin with a simple one hidden layer network as in Fig.~\ref{fig:l1} (plus the ramp layer, as discussed) with $n=128$ neurons, a training box defined by $k_{\rm max}=10$ and a training (validation) set consisting of 70\% (30\%) of zeroth cohomologies within the training box. Training is carried out within about a minute on a single CPU and leads to a success rate of $0.83$. With further fine-tuning and increasing the number of neurons to $n=256$ we are able to push the success rate to $0.85$. While this sounds respectable, it has to be kept in mind that the percentage of vanishing cohomologies within the training box is 73\%. While most of these vanishing cohomologies are correctly reproduced by the net, basically as a direct consequence of including the ramp layer, and, hence, account for a large share of the success rate, only less than half of the non-zero cohomologies are predicted correctly. Similar results are obtained for networks with two hidden layers, as in Fig.~\ref{fig:l2} (again, with a ramp layer added) applied to the zeroth cohomology, as well as for one and two layer networks applied to the first cohomology. It is worth stressing that, just as for two-folds, all success rates dramatically drop outside the training box. 

An improvement on predicting the positive cohomology values can be achieved by modifying the training set to
\begin{equation}
 \left\{{\bf k}\;\rightarrow h^q({\cal O}_X({\bf k}))^{1/p}\;\,|\, |k_i|\leq k_{\rm max},\; h^q({\cal O}_X({\bf k}))>0\right\}\; . \eqlabel{newset}
\end{equation} 
This means we train only on the strictly positive cohomologies and we reduce the large hierarchies in the cohomology dimensions by taking the power $1/p$, where $p=6$, say. With this modification, we can train a single layer network with $n=128$ to correctly reproduce 80\% of the non-zero cohomologies $h^0$, within the training box $k_{\rm max}=10$. Admittedly, this method does to some degree rely on the underlying mathematical structure - basically on a vanishing theorem for the cohomology in question - and, therefore, goes beyond simple function learning.\\[2mm]

To confirm some of these results, we consider a second CICY manifold, the tetra-quadric, defined by the configuration matrix
\begin{equation}
 X\in\left[\begin{array}{c|c}\mathbb{P}^1&2\\\mathbb{P}^1&2\\\mathbb{P}^1&2\\\mathbb{P}^1&2\end{array}\right]\; .
\end{equation}
The rank of the Picard group is four and line bundles ${\cal O}_X({\bf k})$ are labelled by an integer vector ${\bf k}=(k_1,k_2,k_3,k_4)$ with four components.  

A single layer network, as in Fig.~\ref{fig:l1} (with an added ramp layer) and $n=128$ neurons  leads to a superficially impressive success rate of $88\%$ for the zeroth cohomology, within the training box defined by $k_{\rm max}\leq 5$.  However, similar to the bi-cubic case, it turns out that $89\%$ of cohomologies within the training box vanish. This is essentially what the network reproduces correctly, but few of the non-zero cohomologies are predicted accurately. By increasing the number of neurons and/or going to two-layer networks the situation can be improved somewhat, but we have not been able to achieve success rates significantly larger than $0.9$. Similar statements hold for the first cohomology. 

As for the bi-cubic, we can improve the performance on the positive cohomology by modifying the training set to \eqref{newset}. In this case, a single layer network with $n=32$ can achieve a success rate of $0.8$ for the zeroth cohomology and within the training box $k_{\rm max}=5$. 

\subsection{Remarks}
Simple function learning with fully-connected networks of line bundle cohomology appears to work well and can lead to accurate predictions for 80\% or 90\%, sometimes more, of cohomologies within the training box. These results can usually be achieved with a training set which consists of a fraction of the cohomology values in the training box. The network, once trained, is able to predict efficiently, in contrast with the lengthy and time-consuming calculations underlying the various algorithmic methods. 

Despite these advantages, this approach to machine learning of line bundle cohomology is of limited practical use, due to a number of shortcomings.
\begin{itemize}
\item Since cohomology values represent dimensions of vector spaces, the appropriate measure of training success is the fraction of values correctly reproduced after rounding to the nearest integer. Especially for large cohomology dimensions this measure is ambitious and requires extremely accurately trained networks. This affects three-fold cohomology which increases cubicly more readily than cohomology on surfaces which rises quadratically.
\item Even well-trained networks may not be of much practical use for many applications. For example, a network with a success rate of $0.95$ may be applied to string model building. If each model involves $20$ line bundles (not an untypical value) the network will predict the correct spectrum for only about a third of the models.
\item The network is unable to reliably predict cohomology dimensions outside the training box.  This means access to large cohomologies where algorithmic methods are often too slow is not improved.
\item The functional form of cohomology dimensions - vanishing on a large portion of the Picard lattice and rising as a polynomial elsewhere - is not well-suited for simple function learning with logistic sigmoid functions, as our discussion of cohomology on three-folds shows. More generally, approximating a piecewise polynomial function by a superposition of transcendental functions does not appear to be well-suited.
\item The trained network provides little indication of the underlying mathematical structure - the sub-division of the Picard lattice into regions and the polynomial behaviour.
\end{itemize}
All this suggests a different approach to machine learning of line bundle cohomology which incorporates information about the mathematical structure of the problem. This is what we turn to now.

\section{Learning cohomology formulae}\label{learnform}
As discussed earlier, there are strong indications that cohomology dimensions can be described in terms of a relatively simple mathematical structure. The Picard lattice is divided into regions - frequently, but not always cones - in each of which the cohomology dimensions are given by a polynomial with degree equal to the complex dimension of the manifold. This behaviour has been empirically observed for a number of examples~\cite{Constantin:2018hvl,paperex}, including surfaces and three-folds, and it has been proven for classes of surfaces~\cite{papermath} as well as for some three-folds, such as the bi-cubic~\cite{Constantin:2018hvl}. Our hypothesis here is simply that this structure is fairly general and in particular applies to all examples we will study. Where currently no mathematical proof is available, further evidence for the hypothesis will in fact be provided by the approach to machine learning discussed in this section.

\subsection{Network structure}
\begin{figure}[h!!!]
\begin{center}
\includegraphics[width=15cm]{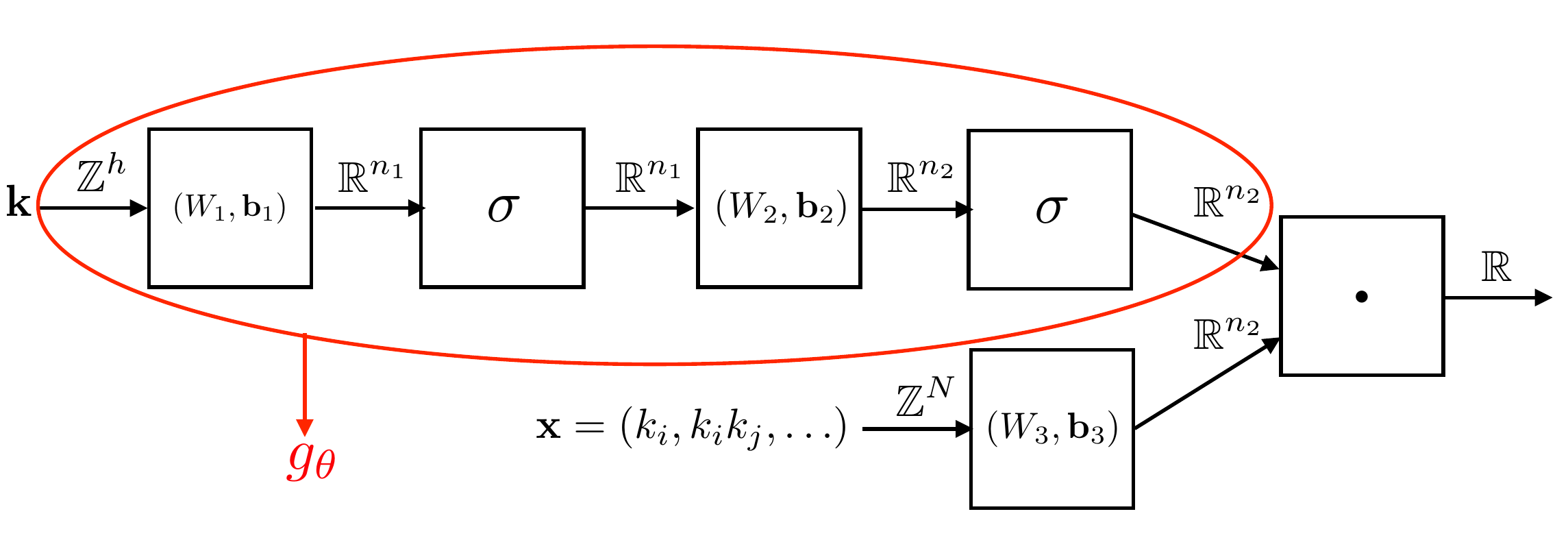}
\caption{\sf Structure of the network to learn piecewise polynomial cohomology formulae. The dot in the element on the right indicates a dot product, taken between the two input vectors.}
\label{fig:fl}
\end{center}
\vskip -5mm
\end{figure}

The structure of the network we will be using is indicated in Fig.~\ref{fig:fl}. 

Let us discuss how this network relates to the underlying mathematical structure of line bundle cohomology. The training data used for this network is of the form~\eqref{tdlin}, that is we are adding the monomials $k_ik_j$ and, for three-folds, also the monomials $k_ik_jk_l$ to the input vector $k_i$, so that the problem becomes effectively piecewise linear. 

The upper part of the network in Fig.~\ref{fig:fl}, denoted by $g_\theta$, where $\theta=(W_1,{\bf b}_1,W_2,{\bf b}_2)$, is a two-layer network with $n_1$ and $n_2$ neurons in each layer, whose input is simply the vector ${\bf k}$. We can think of the output, $g_\theta({\bf k})$ as a ``binary code" which indicates the region within which the vector ${\bf k}$ resides. In other words, this upper branch of the network detects the various regions in the Picard lattice. The width $(n_1,n_2)$ of this network should be of the order of the expected number of regions.  The underlying assumption is that the regions are given as intersections of half-spaces, which is the case for most examples studied to date.

The lower part of the network is a simple linear layer with $n_2$ neurons and weights/biases $(W_3,{\bf b}_3)$ which represent the coefficients of the various polynomial equations. By multiplying these weights into the monomial input vector ${\bf x}=(k_i,k_ik_j,\ldots)$ this layer outputs a vector which consists of the values of $n_2$ polynomials in ${\bf k}$. Dotting this vector into the ``binary vector" from the upper network selects the specific linear combination of polynomials associated to  the region in which ${\bf k}$ resides.\\[2mm]
Suppose the above network has been trained leading to the values $\bar{\theta}$ and $(\bar{W}_3,\bar{\bf b}_3)$ of the weights and biases. Of course this network can then be used to predict individual cohomology dimensions, just as the conceptually simpler networks studied in the previous section. However, our main purpose here is more ambitious. 

Suppose that for each vector ${\bf k}$ inside the training box we compute, after some suitable network surgery to extract $g_{\bar{\theta}}$ and $(\bar{W}_3,\bar{\bf b}_3)$ from the trained network, the vector
\begin{equation}
 {\bf a}({\bf k}):=\left(g_{\bar{\theta}}({\bf k})\cdot \bar{\bf b}_3\,,\;g_{\bar{\theta}}({\bf k})\cdot\bar{W}_3\right)\; .
\end{equation} 
Intuitively, we should think of ${\bf a}({\bf k})$ as the vector of coefficients in the polynomial which describes the cohomology of ${\cal O}({\bf k})$. This intuition leads us to  identify two vectors ${\bf k}$ and ${\bf k}'$ as being described by the same polynomial and, hence, as belonging to the same region in the Picard lattice by using the following condition.
\begin{equation}
 {\bf k},\;{\bf k}'\mbox{ in the same region}\quad\leftrightarrow\quad|{\bf a}({\bf k})-{\bf a}({\bf k}')|<\epsilon \eqlabel{kkeq}
\end{equation}
Here, $\epsilon>0$ is suitably small. In this way, we can identify the regions in the Picard lattice as point sets within the training box. 

Once this is accomplished, we are almost finished. For each point set, we can select a number of points - preferably in the interior of the set - and their cohomology values and fit to this subset a polynomial of the appropriate degree.  In short, once the regions have been found the polynomials are easily determined. Turning this around, with the polynomials at hand, their range of validity within the training box is easily established. This can be used to ``clean up" the regions which are typically not precise, particularly along the boundaries, when determined from the network using Eq.~\eqref{kkeq}. With the precise regions known as point sets, conventional algorithms can then be used to determine their defining inequalities. Altogether, this provides a method to generate a conjecture for a cohomology formula, similar to the ones in Eqs.~\eqref{dP1h0}, \eqref{H0:bicubic}, \eqref{H1:bicubic}, from machine learning. \\[2mm]
To summarise, our algorithm has the following main steps.
\begin{itemize}
\item[{\bf (1)}] Train the network in Fig.~\ref{fig:fl} with line bundle cohomology data for a manifold, given in the form~\eqref{tdlin}. The width $(n_1,n_2)$ of the network is varied and fixed by optimising the success rate.
\item[{\bf (2)}] From the trained network, find the regions in the Picard lattice by using the relation~\eqref{kkeq}.
\item[{\bf (3)}] For each so-obtained region, find the corresponding polynomial by a simple fit to points in the interior of the region.
\item[{\bf (4)}] Use these polynomials to determine the exact regions of their validity.
\item[{\bf (5)}] Determine the bounding inequalities of these exact regions, using standard algorithms.
\end{itemize}
A simplifying assumption underlying this algorithm is that there is indeed only one polynomial per region which describes the cohomology dimensions. Sometimes, it happens that the cohomology in a given region shows an alternating even/odd pattern, so that two polynomials are required (higher order patterns can also arise). This has been observed in Ref.~\cite{Klaewer:2018sfl} and, in the case of surfaces, this behaviour can be traced back to the ceiling function in the master formula~\eqref{master} (see Ref.~\cite{paperex} for details). For the present section, we have selected examples which do not show such an alternating behaviour and we will now demonstrate that the algorithm works successfully for such examples.

\subsection{Surface examples}
We begin with the zeroth cohomology on the del Pezzo surface $\dps{1}$, in part to illustrate the algorithm in a simple case. Recall that the rank of the Picard group is two and line bundles ${\cal O}_X({\bf k})$ are parametrised by a two-dimensional integer vector ${\bf k}=(k_0,k_1)$.

We use a training box $k_{\rm max}=15$ and train on the entire set of cohomologies within this box. (There is no need for sampling or validation - the trained network is validated by leading to a sensible formula for cohomology.) We do not know, a priori, how many neurons $(n_1,n_2)$ the network in Fig.~\ref{fig:fl} should have. For this reason, we run over a number of sizes, $(n_1,n_2)\in\{(1,1),(1,2),(2,1),(2,2),(2,3),(3,2),(3,3)\}$ and choose the minimal configuration with a high success rate. It turns out that $(n_1,n_2)=(2,2)$ is adequate.

\begin{figure}[h!!!]
\begin{center}
\includegraphics[width=8cm]{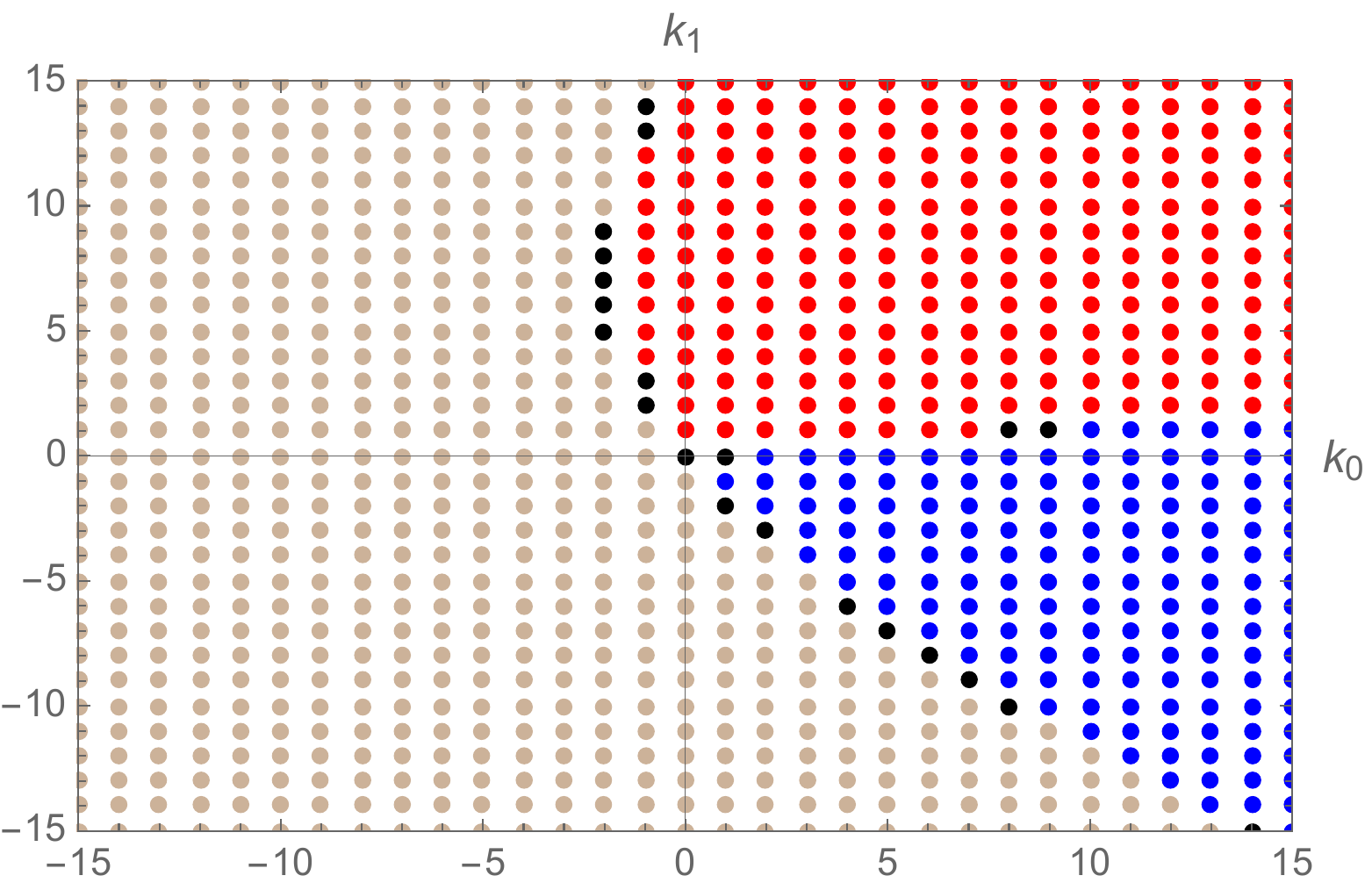}
\includegraphics[width=7.9cm]{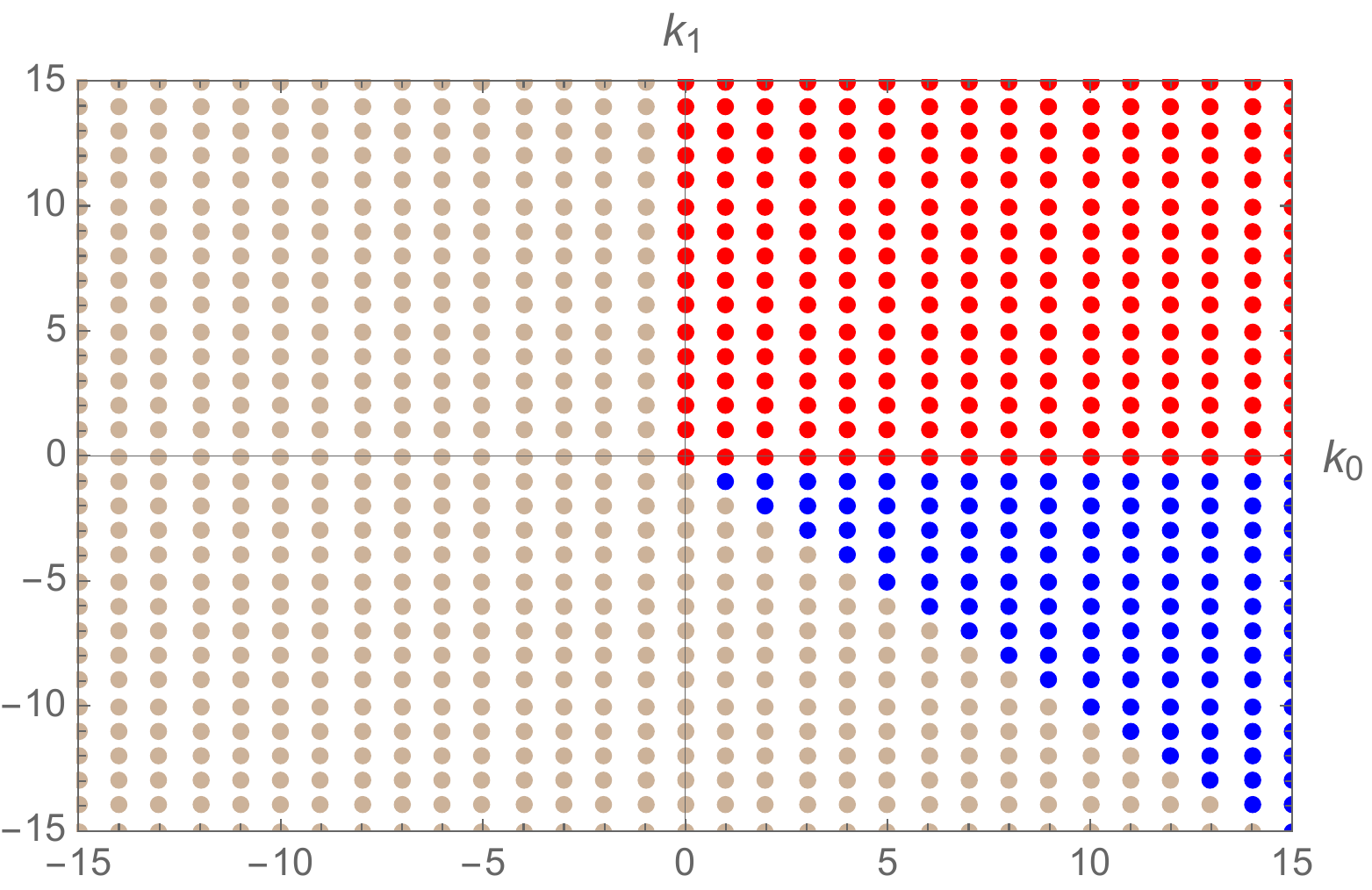}
\caption{\sf Regions in the Picard lattice for the zeroth cohomology of line bundles on $\dps{1}$, determined from the trained network in Fig.~\ref{fig:fl} for $(n_1,n_2)=(2,2)$ and a training box $k_{\rm max}=15$. The figure on the left shows the regions obtained after step (2) of the algorithm. (Black points indicate spurious regions which should be ignored.) The figure on the right shows the ``cleaned-up" regions obtained after step (4) of the algorithm.}
\label{fig:dP1reg}
\end{center}
\vskip -5mm
\end{figure}

With this network trained, we perform step (2) of the algorithm, by using the relation~\eqref{kkeq}. In this way, we arrive at the plot on the left-hand-side of Fig.~\ref{fig:dP1reg} which shows the three regions which have been identified. 
As is evident from the plot, the boundaries of the regions are not necessarily exact. Nevertheless, we can choose, in each region, a small number of points away from the boundaries and fit a quadric in $k_i$ to the cohomology values of these points. The yellow, green and blue regions clearly correspond to the three rows in the formula~\eqref{dP1h0} and the three polynomials we find from the fit are precisely those given in Eq.~\eqref{dP1h0}. This completes the third step of the algorithm. 

With the polynomials at hand, we can now carry out step (4) and find the precise regions in the Picard lattice where they reproduce the correct cohomology dimension. This leads to the plot on the right-hand-side of Fig.~\ref{fig:dP1reg}. In the final step (5), we can then work out the inequalities which describe those exact regions, by fitting hyperplane equations to the boundary points. The polynomials found earlier together with these inequalities then reproduce the entire analytic formula~\eqref{dP1h0} for the zeroth cohomology on $\dps{1}$.\\[2mm]
Our next example is for the zeroth cohomology on the del Pezzo surface $\dps{2}$. The rank of the Picard group is three and line bundles ${\cal O}_X({\bf k})$ are parametrised by integer vectors ${\bf k}=(k_0,k_1,k_2)$. We are using a training box of size $k_{\rm max}=15$ and, after scanning over a range of networks widths $n_1,n_2\in\{1,\ldots 10\}$ settle for a configuration $(n_1,n_2)=(8,8)$.
\begin{figure}[h!!!]
\begin{center}
\includegraphics[width=9cm]{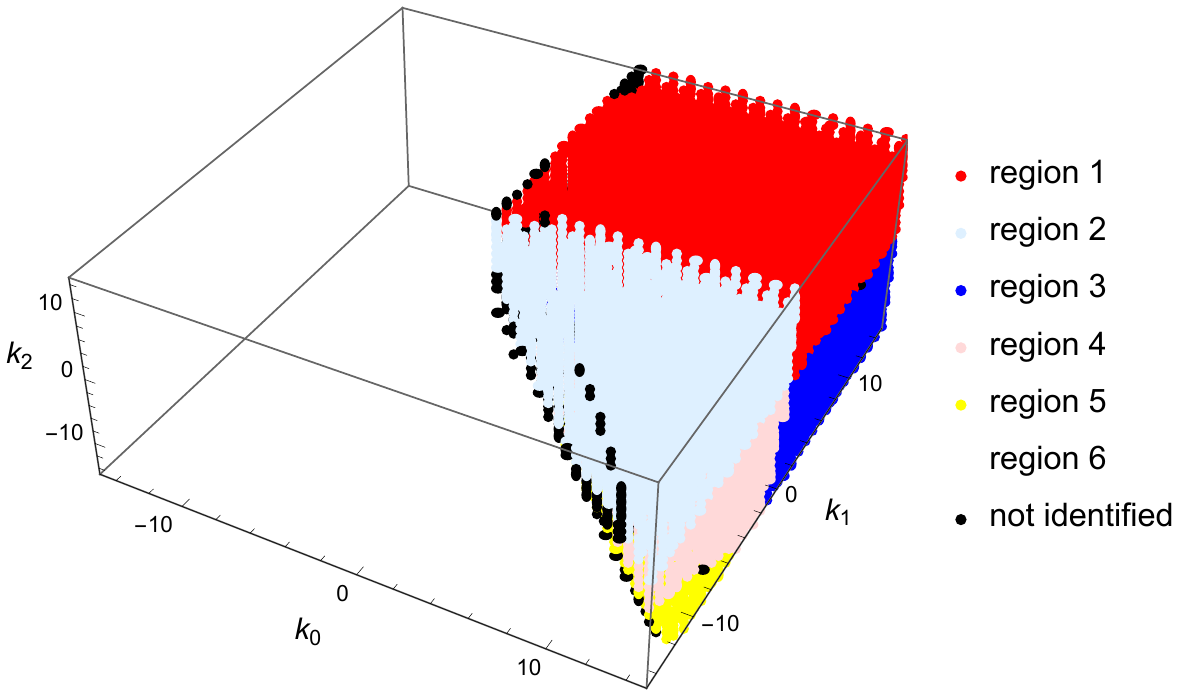}
\includegraphics[width=7cm]{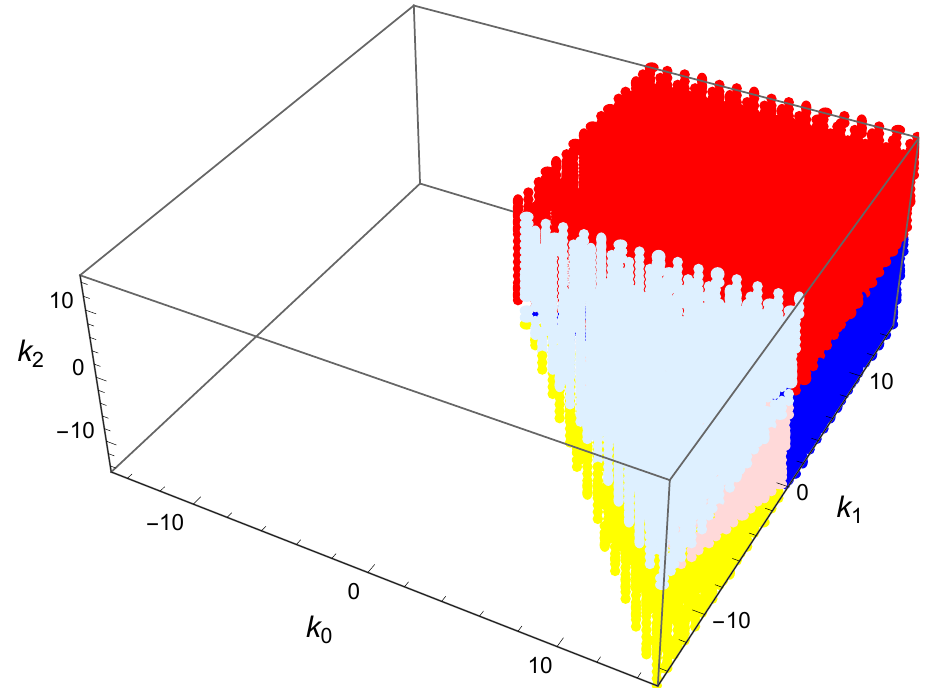}
\caption{\sf Regions in the Picard lattice for the zeroth cohomology of $\dps{2}$, determined from the trained network in Fig.~\ref{fig:fl} for $(n_1,n_2)=(8,8)$ and a training box of size $k_{\rm max}=15$. The figure on the left shows the regions obtained after step (2) of the algorithm. The figure on the right shows the ``cleaned-up" regions obtained after step (4) of the algorithm. The legend labels the regions as in Eqs.~\eqref{dP2eq}, \eqref{dP2reg}}
\label{fig:dP2reg}
\end{center}
\vskip -8mm
\end{figure}

Training the network within the training box $k_{\rm max}=15$ and performing the identification from Eq.~\eqref{kkeq} (step (2) of the algorithm) leads to the point sets plotted on the left-hand-side of Fig.~\ref{fig:dP2reg}. A polynomial fit to each of the six regions (step (3) of the algorithm) leads to
\begin{equation}\eqlabel{dP2eq}
h^0({\cal O}_{\dps{2}}({\bf k})) = 
\begin{cases}
1+\frac{3}{2}k_0+\frac{1}{2}k_0^2+\frac{1}{2}k_1-\frac{1}{2}k_1^2+\frac{1}{2}k_2-\frac{1}{2}k_2^2 & \textrm{in region 1} \,, \\
1+2k_0+k_0^2+k_1+k_0k_1+k_2+k_0k_2+k_1k_2 & \textrm{in region 2} \,, \\
1+\frac{3}{2}k_0+\frac{1}{2}k_0^2+\frac{1}{2}k_2-\frac{1}{2}k_2^2 & \textrm{in region 3} \,, \\
1+\frac{3}{2}k_0+\frac{1}{2}k_0^2+\frac{1}{2}k_1-\frac{1}{2}k_1^2 & \textrm{in region 4} \,, \\
1+\frac{3}{2}k_0+\frac{1}{2}k_0^2 & \textrm{in region 5} \,. \\
0& \textrm{in region 6} \,. \\
\end{cases}
\end{equation}
Using these equations to determine the exact regions (step (4)) leads to the plot on the right-hand-side of Fig.~\ref{fig:dP2reg}. In step (5), we then determine the inequalities which describe those regions. They are given by
\begin{equation}\eqlabel{dP2reg}
\begin{tabular}{ccccccc}
Region 1: & $-k_1 \geq 0$	& $-k_2 \geq 0$& $k_0+k_1+k_2 \geq 0$ 	& 				& 				& \\
Region 2: & 			&  			& $k_0+k_1+k_2 < 0$& $k_0+k_1 \geq 0$	& $k_0+k_2 \geq 0$ 	& \\
Region 3: & $-k_1 < 0$& $-k_2 \geq 0$& 					& 			 	& $k_0+k_2 \geq 0$ 	& \\
Region 4: & $-k_1 \geq 0$	& $-k_2 < 0$& 					& 			 	& $k_0+k_2 \geq 0$ 	& \\
Region 5: & $-k_1 < 0$& $-k_2 < 0$	& 					& 			 	&  				& $k_0 \geq 0$ \\
Region 6: & otherwise&&&&\\
\end{tabular}
\end{equation}
In summary, the network has learned the formula for the dimensions of the zeroth line bundle cohomology on $\dps{2}$. By applying the master formula~\eqref{master} to $\dps{2}$, it can be shown that the above result is indeed correct on the entire Picard lattice.  The explicit proof can be found in Ref.~\cite{paperex}.\\[2mm]
We would like to analyse two further surface examples for which cohomology formulae are not yet known. They are CI manifolds defined by the configuration matrices
\begin{equation}
 X\in\left[\begin{array}{c|c}\mathbb{P}^1&2\\\mathbb{P}^2&3\end{array}\right]\;,\qquad
 Y\in\left[\begin{array}{c|c}\mathbb{P}^1&3\\\mathbb{P}^2&4\end{array}\right]\; . \eqlabel{K3}
\end{equation} 
The first of these, $X$, is a K3 surface, while $Y$ is a surface of general type with an ample canonical bundle. For both cases, the rank of the Picard lattice is two and line bundles are denoted by ${\cal O}({\bf k})$, where ${\bf k}=(k_1,k_2)$. 
\begin{figure}[h!!!]
\begin{center}
\includegraphics[width=8cm]{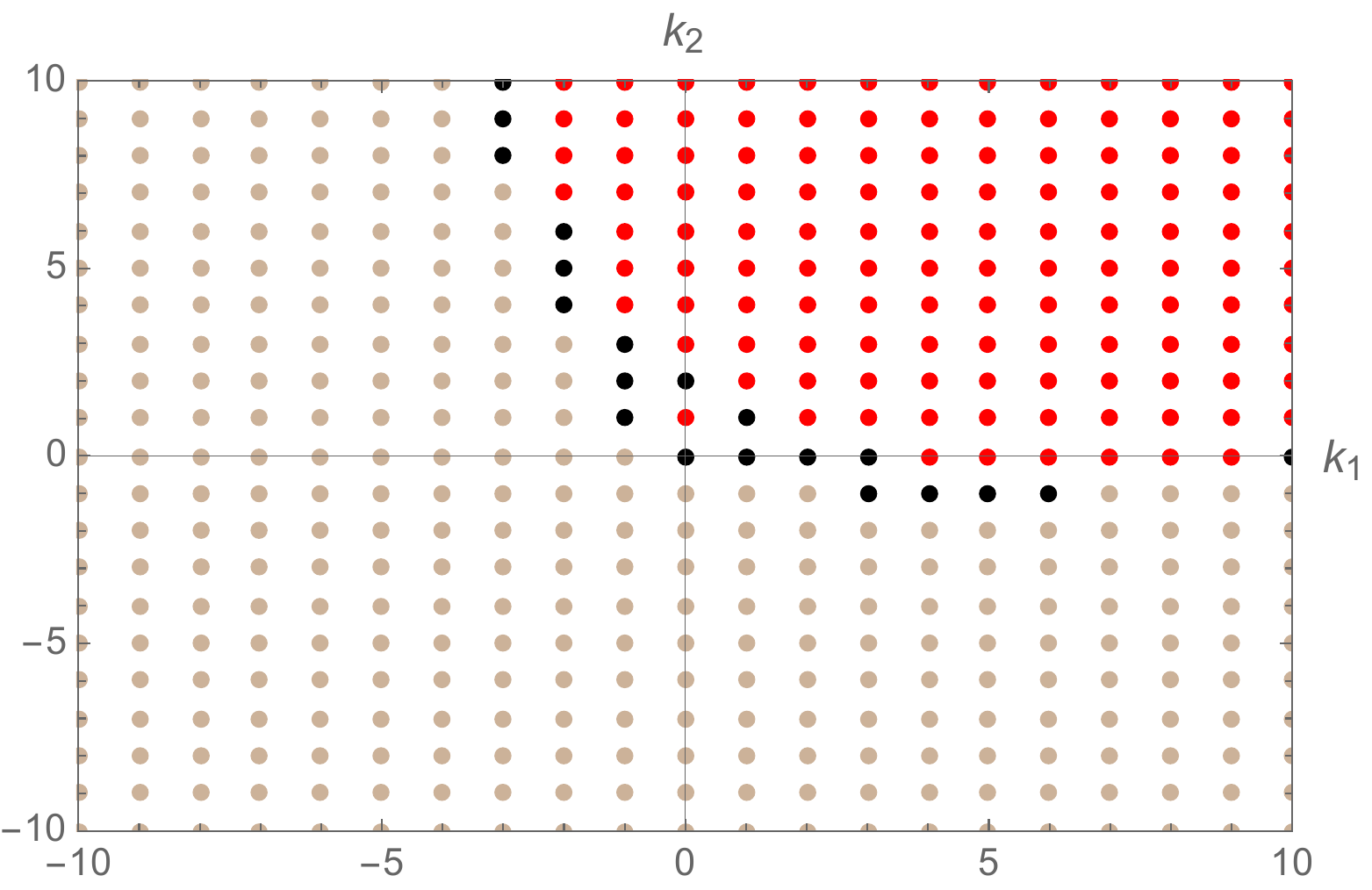}
\includegraphics[width=8cm]{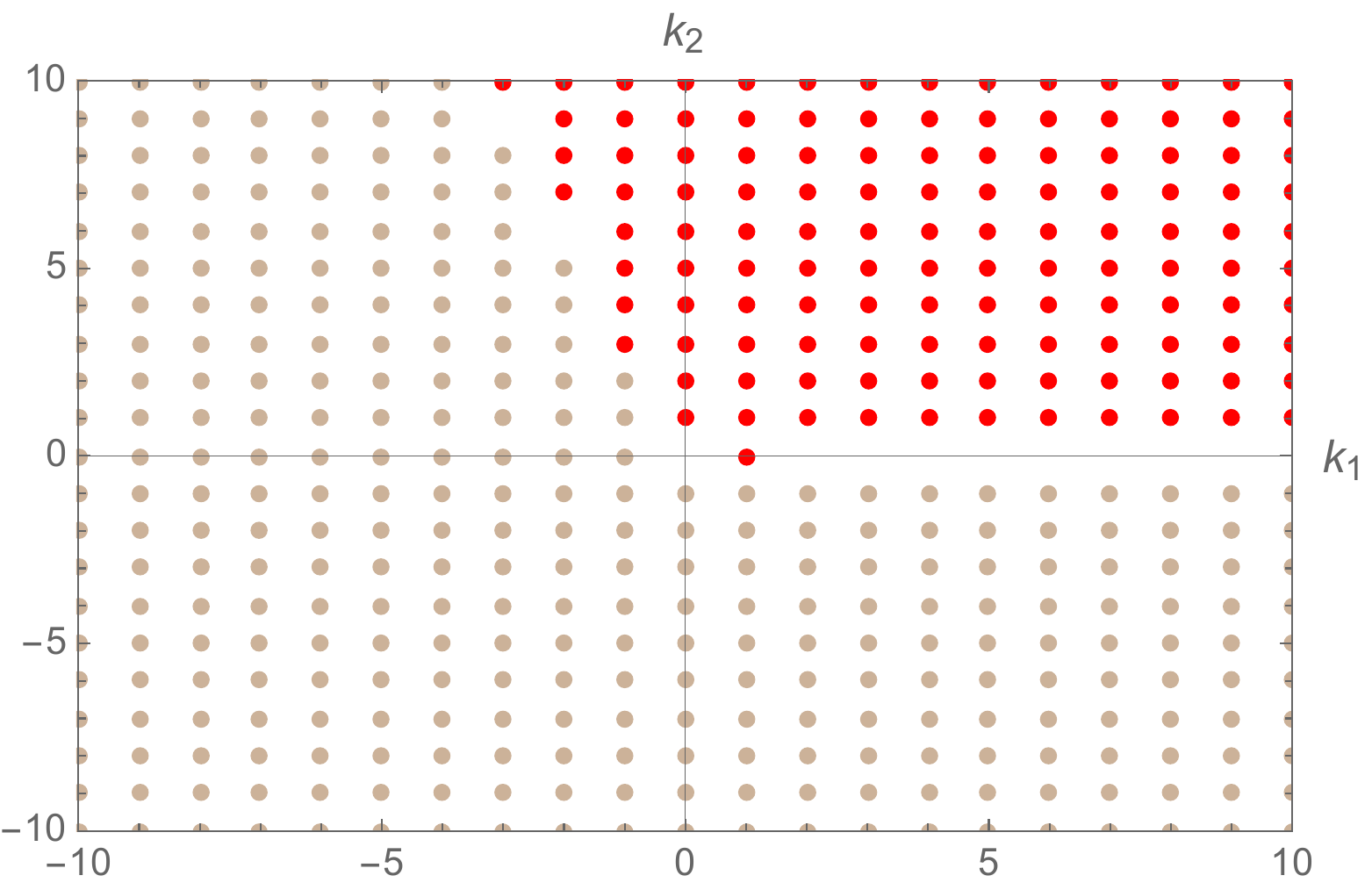}
\caption{\sf Regions in the Picard lattice for the zeroth cohomology of the K3 surface $X$ in Eq.~\eqref{K3}, determined from the trained network in Fig.~\ref{fig:fl} for $(n_1,n_2)=(3,3)$ and a training box of size $k_{\rm max}=10$. The figure on the left shows the regions obtained after step (2) of the algorithm. The figure on the right shows the ``cleaned-up" regions obtained after step (4) of the algorithm.}
\label{fig:K3}
\end{center}
\vskip -8mm
\end{figure}
\begin{figure}[h!!!]
\begin{center}
\includegraphics[width=8cm]{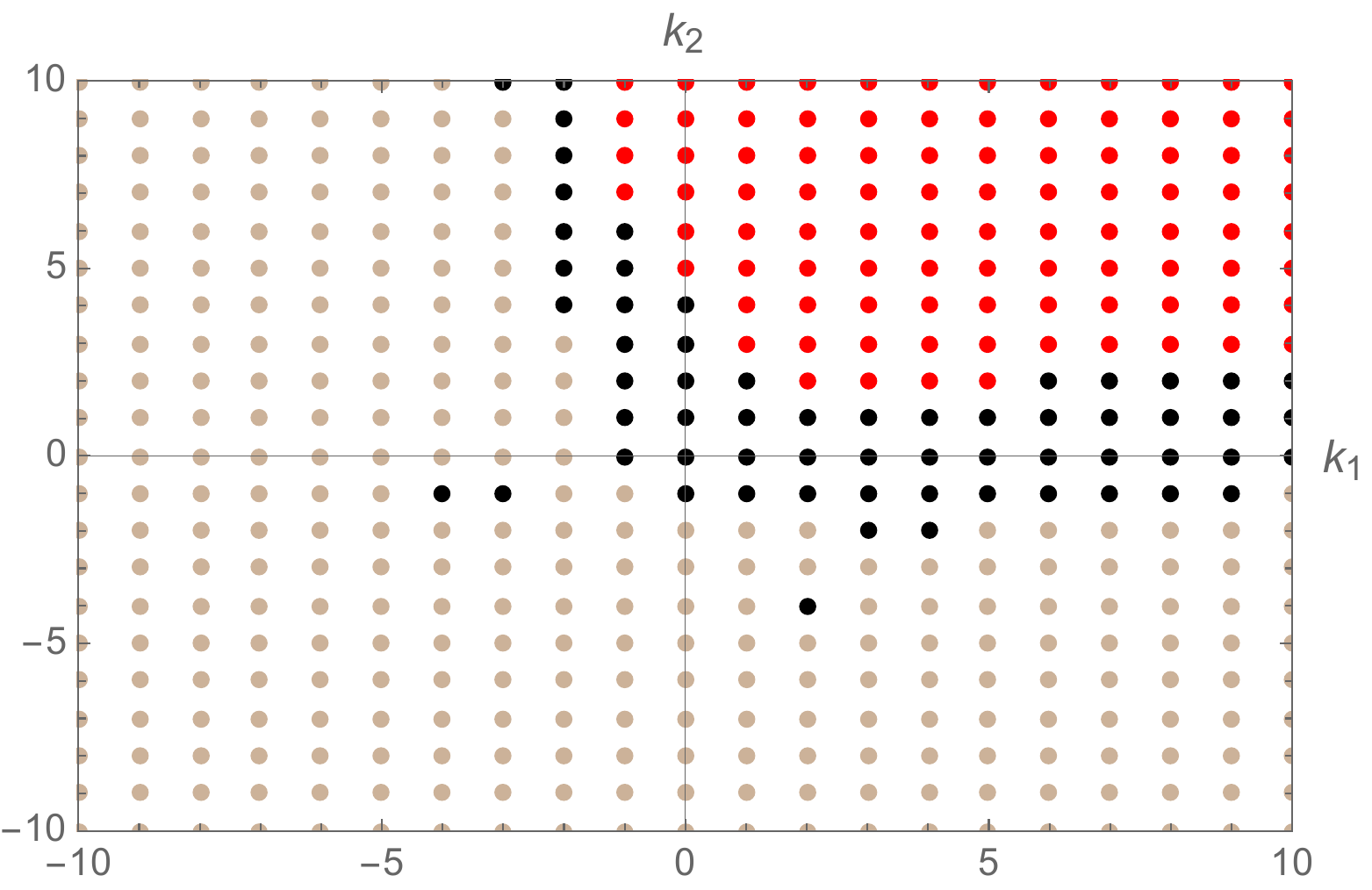}
\includegraphics[width=8cm]{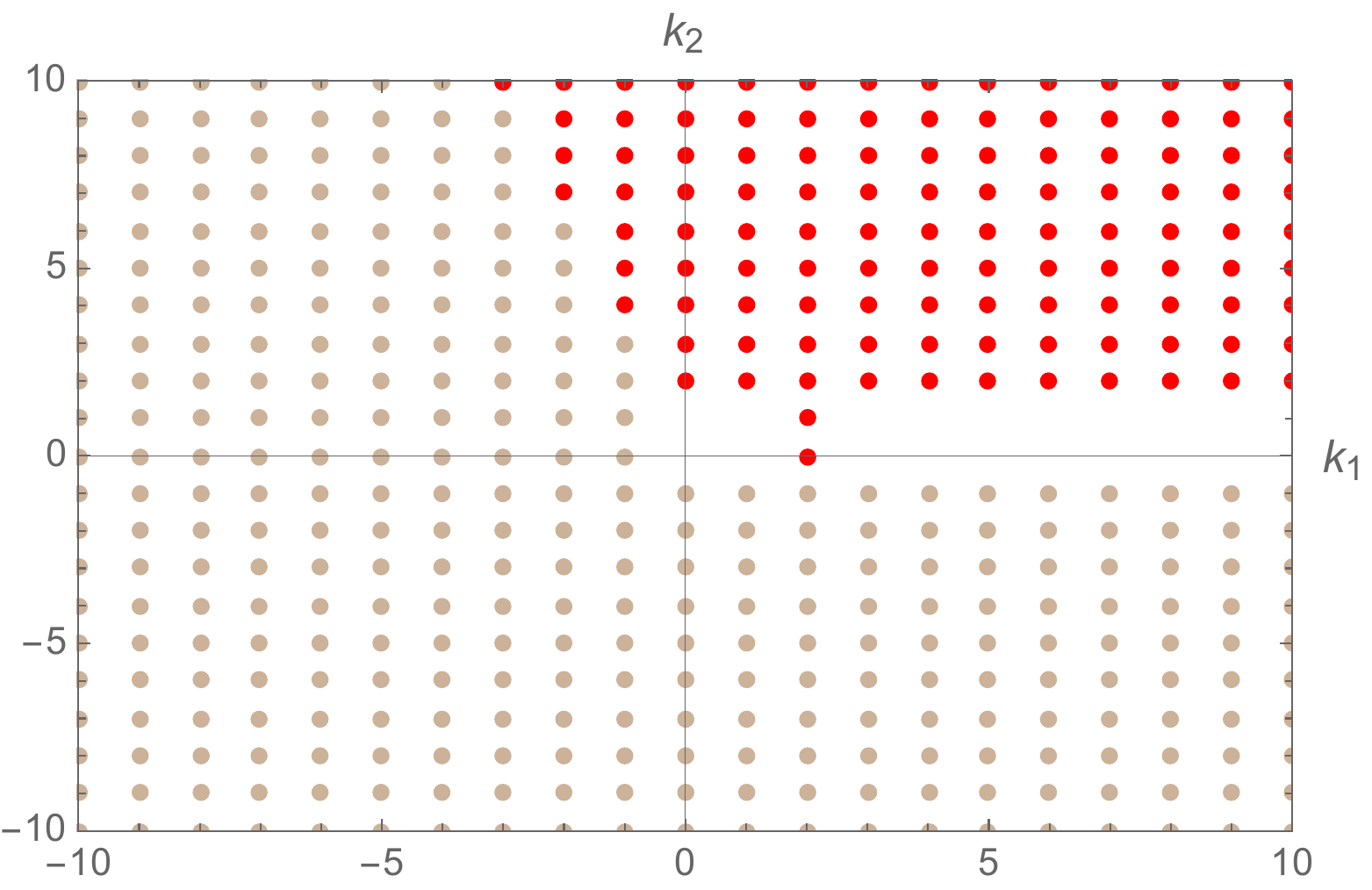}
\caption{\sf Regions in the Picard lattice for the zeroth cohomology of the surface $Y$ of general type in Eq.~\eqref{K3}, determined from the trained network in Fig.~\ref{fig:fl} for $(n_1,n_2)=(3,3)$ and a training box of size $k_{\rm max}=10$. The figure on the left shows the regions obtained after step (2) of the algorithm. The figure on the right shows the ``cleaned-up" regions obtained after step (4) of the algorithm.}
\label{fig:gen}
\end{center}
\vskip -8mm
\end{figure}
The results for the K3 example are shown in Fig.~\ref{fig:K3}. As is evident, the network identifies two large regions. We can also see a new phenomenon emerging which does not arise for manifolds with an ample anti-canonical bundle. There are lower-dimensional regions, in the present case along the positive $k_1$ axis and along the line $3k_1+k_2=0$ for $k_2>0$, which are not identified by the network since they contain too few points within the training box. We can cover those regions by adding a further step to our algorithm. We consider all points in the training box left over after step (5) and then attempt to fit polynomials to these remaining points. This leads to the cohomology formula
\begin{equation}
 h^0({\cal O}_X({\bf k}))=\left\{\begin{array}{ll}2+k_1k_2+k_2^2&\mbox{for}\quad3k_1+k_2>0\mbox{ and }k_2>0\\
                                                                      0&\mbox{for} \quad3k_1+k_2<0\mbox{ or }k_2<0\\
                                                                      k_1+1&\mbox{for}\quad k_1\geq 0\mbox{ and }k_2=0\\
                                                                      -k_1+1&\mbox{for}\quad3k_1+k_2=0\mbox{ and }k_2>0\end{array}\right.\; .
\end{equation}                                                                      
Our second example, for the surface $Y$ of general type leads to a similar structure, as can be seen in Fig.~\ref{fig:gen}. However, compared to the K3 case, one of the one-dimensional regions has disappeared, while the other one is enlarged to width two. (The two red points in this region on the right-hand-side plot in Fig.~\ref{fig:gen} correspond to ``accidental" matching of the cohomology formula.) A polynomial fit to the three regions leads to
\begin{equation}
 h^0({\cal O}_Y({\bf k}))=\left\{\begin{array}{ll}7-2k_1-\frac{7}{2}k_2+4k_1k_2+\frac{3}{2}k_2^2&\mbox{for}\quad k_2\geq 2\mbox{ and }3k_1+k_2\geq 1\\
                                                                        1+k_1+2k_2+2k_1k_2&\mbox{for}\quad k_1\geq 0\mbox{ and }k_2\in\{0,1\}\\
                                                                        0&\mbox{otherwise}
                                                                        \end{array}\right.\; .
\end{equation}                                                                        
To our knowledge this is the first example of a cohomology formula for a surface of general type.

\subsection{Three-fold examples}
\begin{figure}[h!!!]
\begin{center}
\includegraphics[width=8cm]{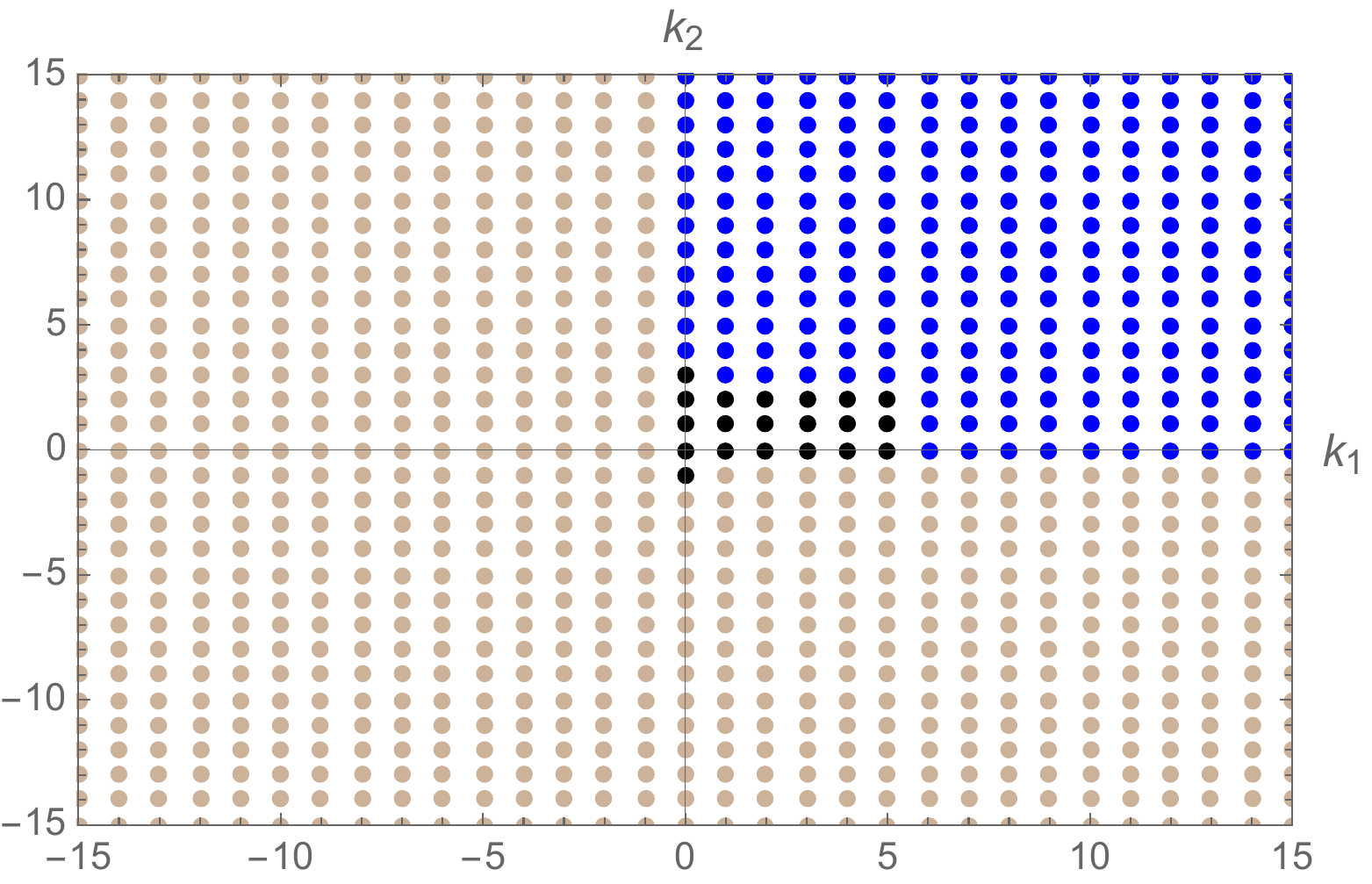}
\includegraphics[width=8cm]{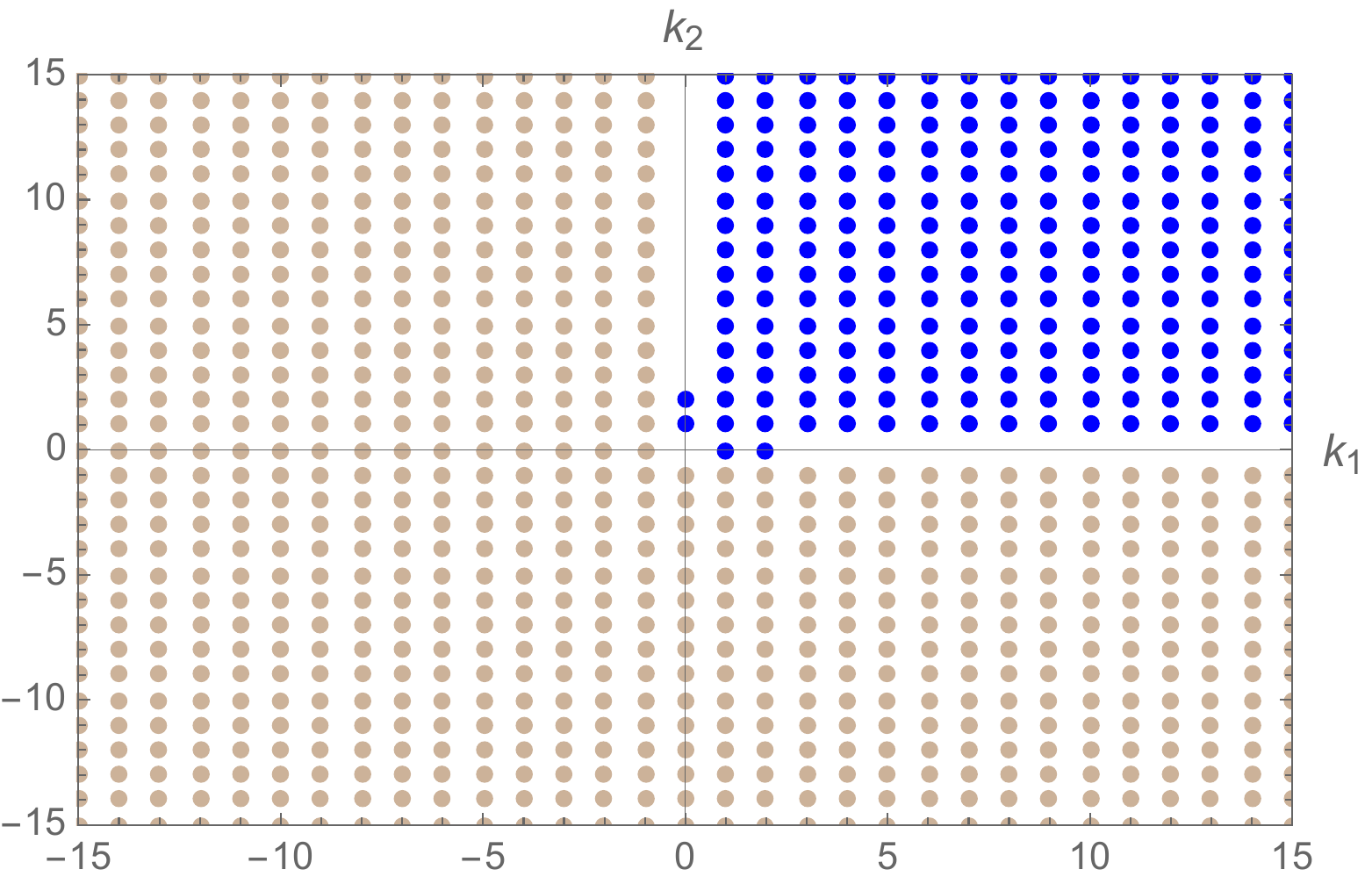}
\caption{\sf Regions in the Picard lattice for the zeroth cohomology of line bundles on the bicubic, determined from the trained network in Fig.~\ref{fig:fl} for $(n_1,n_2)=(3,2)$ and a training box $k_{\rm max}=15$. The figure on the left shows the regions obtained after step (2) of the algorithm. The figure on the right shows the ``cleaned-up" regions obtained after step (4) of the algorithm.}
\label{fig:bicubich0}
\end{center}
\vskip -5mm
\end{figure}

\begin{figure}[h!!!]
\begin{center}
\includegraphics[width=8cm]{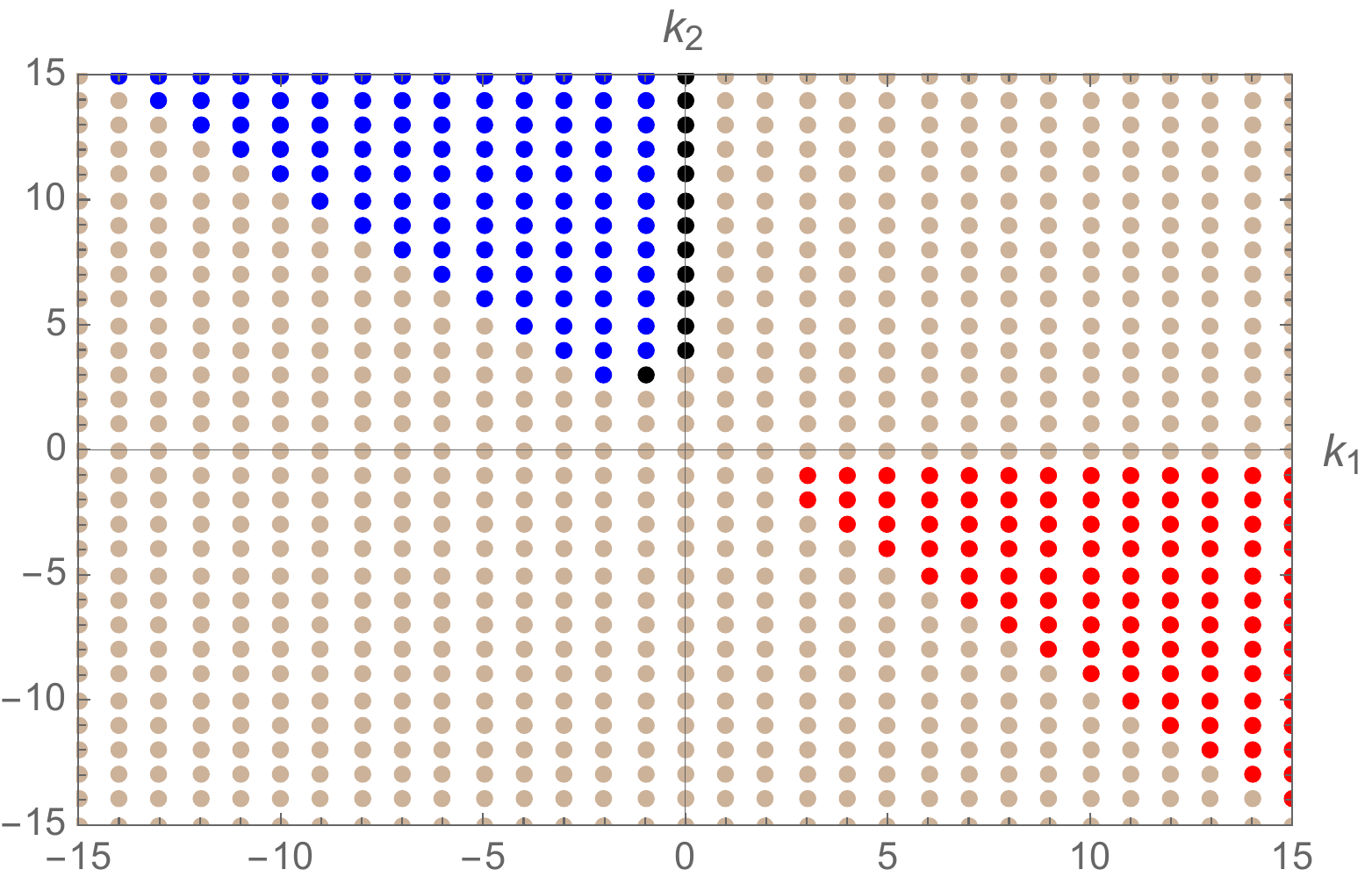}
\includegraphics[width=8cm]{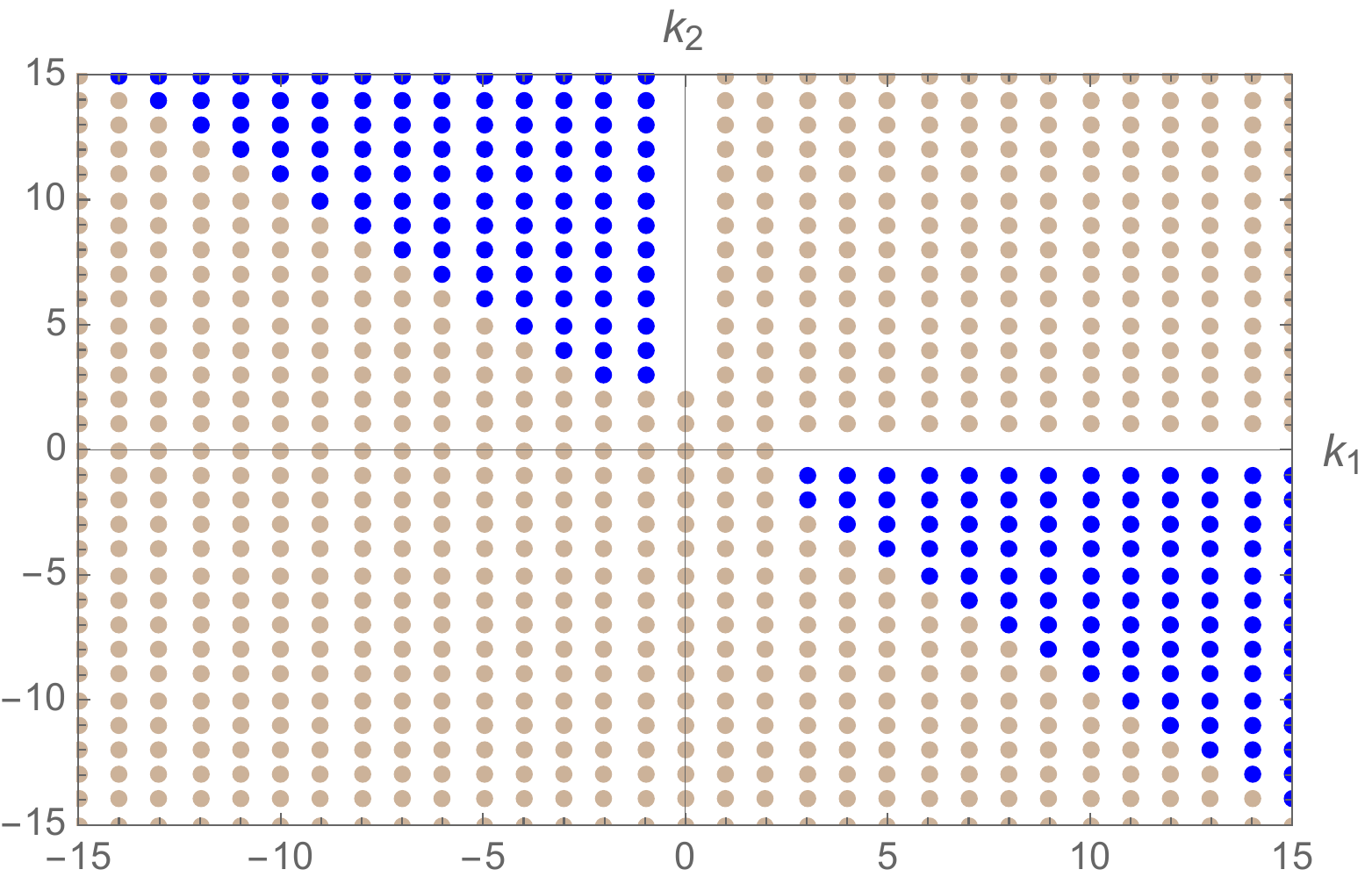}
\caption{\sf Regions in the Picard lattice for the first cohomology of line bundles on the bicubic, determined from the trained network in Fig.~\ref{fig:fl} for $(n_1,n_2)=(4,2)$ and a training box $k_{\rm max}=15$. The figure on the left shows the regions obtained after step (2) of the algorithm. The figure on the right shows the ``cleaned-up" regions obtained after step (4) of the algorithm.}
\label{fig:bicubich1}
\end{center}
\vskip -5mm
\end{figure}

Our first three-fold example is for the bi-cubic CICY defined by the configuration matrix~\eqref{confbicubic}. We recall that line bundles ${\cal O}_X({\bf k})$ are labelled by two-dimensional integer vectors ${\bf k}=(k_1,k_2)$.  We consider both the zeroth and first cohomology, using networks of size $(n_1,n_2)=(3,2)$ and $(n_1,n_2)=(4,2)$, respectively, and a training box $k_{\rm max}=15$. The results for the zeroth and first cohomology are shown in Figs.~\ref{fig:bicubich0} and \ref{fig:bicubich1}, respectively.
Fitting polynomials and finding the equations for the regions directly leads to the formula for $h^0$ given by the second and third row of Eq.~\eqref{H0:bicubic} and to the formula for $h^1$ given by the second and third row of Eq.~\eqref{H1:bicubic}. 

As for the previous K3 example, there are left-over one-dimensional regions both for the zeroth and first cohomology, as can be seen from the right-hand-side plots in Figs.~\ref{fig:bicubich0} and \ref{fig:bicubich1}. Fitting polynomials to these one-dimensional regions leads to the first rows in Eq.~\eqref{H0:bicubic} and \eqref{H1:bicubic} (and their counterparts obtained by exchanging $k_1$ and $k_2$).
In this way, we can reproduce Eqs.~\eqref{H0:bicubic} and \eqref{H1:bicubic} entirely.\\[2mm]
Our final example is for the co-dimension two CICY in the ambient space $\mathbb{P}^1\times\mathbb{P}^4$ defined by the configuration matrix
\begin{equation}
 X\in\left[\begin{array}{c|cc}\mathbb{P}^1&0&2\\\mathbb{P}^4&4&1\end{array}\right]\; . \eqlabel{conf7888}
\end{equation} 
Line bundles ${\cal O}({\bf k})$ on this space are labelled by a two-dimensional integer vector ${\bf k}=(k_1,k_2)$ and we are interested in the zeroth cohomology. At the time this work has been carried out, the formula for the dimension of this cohomology was not known. It has recently been found in Ref.~\cite{Larfors:2019sie}.

\begin{figure}[h!!!]
\begin{center}
\includegraphics[width=8cm]{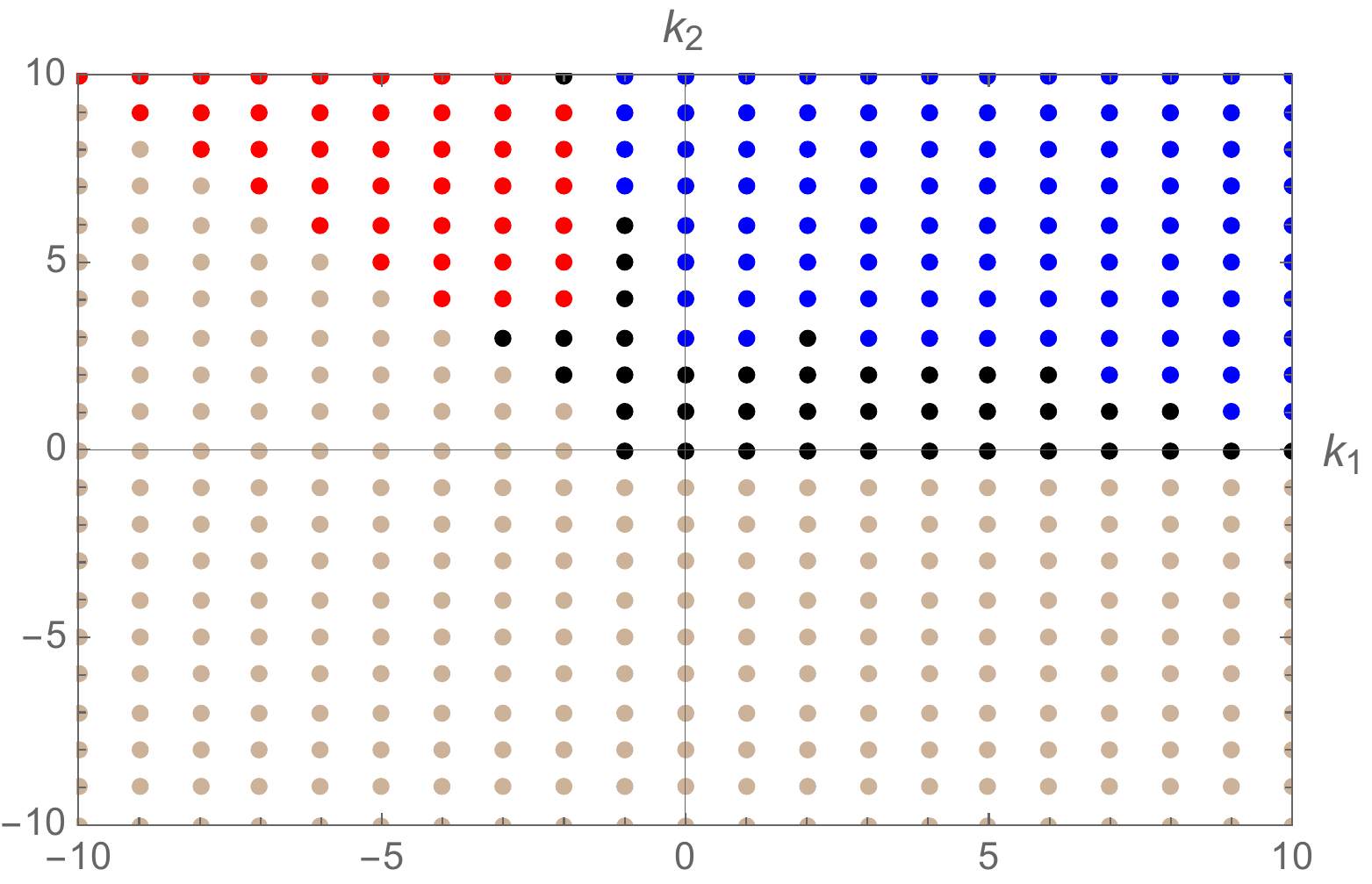}
\includegraphics[width=8cm]{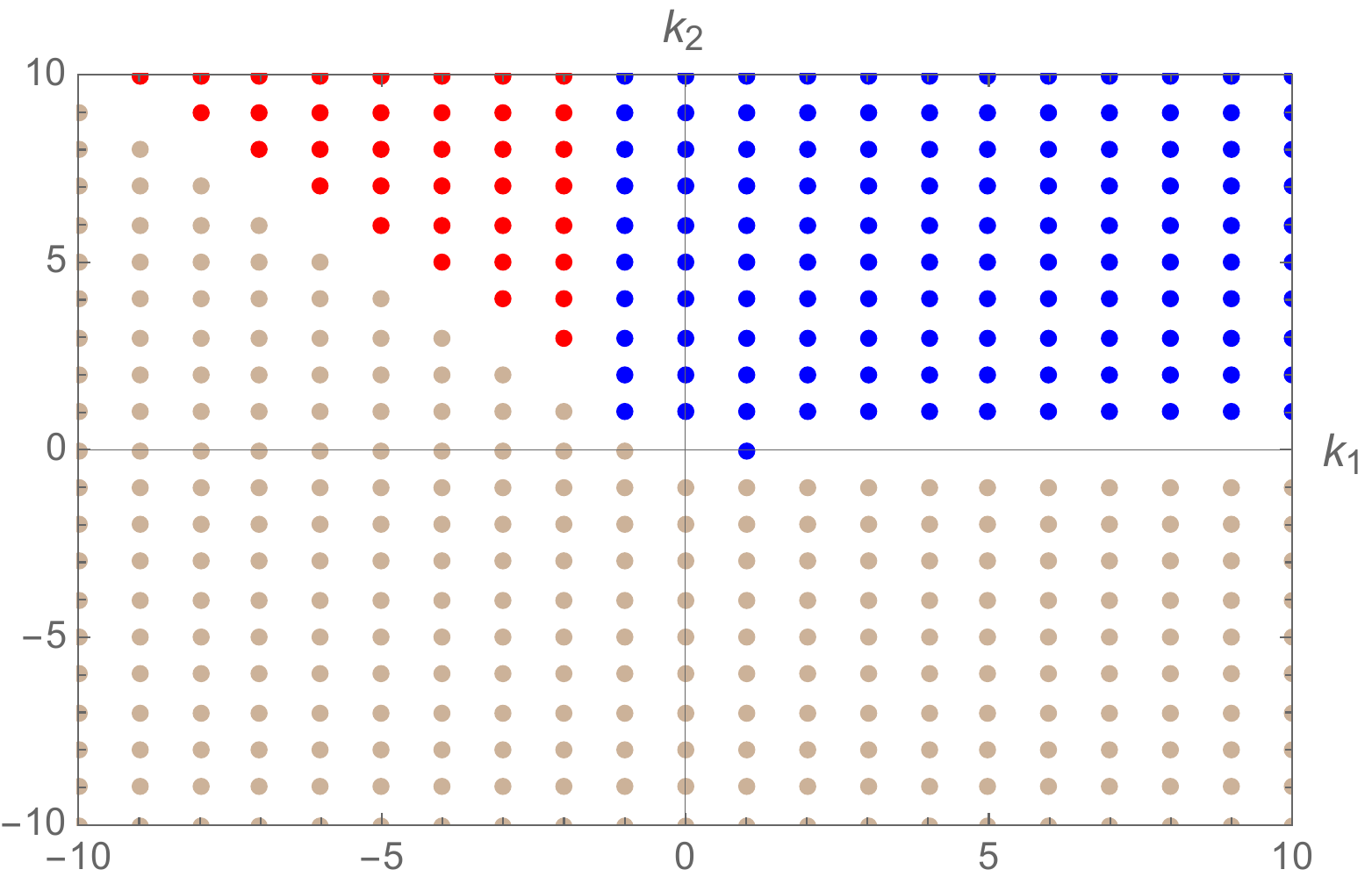}
\caption{\sf Regions in the Picard lattice for the zeroth cohomology of line bundles on the CICY~\eqref{conf7888} determined from the trained network in Fig.~\ref{fig:fl} for $(n_1,n_2)=(4,2)$ and a training box $k_{\rm max}=10$. The figure on the left shows the regions obtained after step (2) of the algorithm. (The black dots correspond to spurious regions which are discarded.) The figure on the right shows the ``cleaned-up" regions obtained after step (4) of the algorithm.}
\label{fig:7888}
\end{center}
\vskip -5mm
\end{figure}

Optimisation of a network as in Fig.~\ref{fig:fl} with $(n_1,n_2)=(4,2)$ and training box $k_{\rm max}=10$ produces the regions in the plot on the left-hand-side of Fig.~\ref{fig:7888}.
Fitting cubics to these regions leads to the polynomials in the first three rows of Eq.~\eqref{7888h0}. From these polynomials we determine the exact regions of validity which are shown in the right-hand-side plot in Fig.~\ref{fig:7888}. As for the bi-cubic, there exist two one-dimensional regions with too few points in the training box to be detected by the network. Fitting polynomials to these left-over points then completes the formula which reads
\begin{equation}
 h^0({\cal O}_X({\bf k}))=\left\{\begin{array}{lll} {\rm ind}({\cal O}_X({\bf k}))&&k_1\geq -1\mbox{ and }k_2>0\\
                                                                     {\rm ind}({\cal O}_X({\bf k}))+\frac{2}{3}k_1-\frac {2}{3}k_1^3&&k_1<-1\mbox{ and } k_1+k_2>0\\
                                                                     0&&k_2<0\mbox{ or }k_1+k_2<0\\
                                                                     k_1+1&&k_1\geq 0\mbox{ and } k_2=0\\
                                                                     k_2+1&&k_2>0\mbox{ and } k_1+k_2=0
                                          \end{array}\right.\; , \eqlabel{7888h0}
\end{equation}                                                                                   
where ${\rm ind}({\cal O}_X({\bf k}))=2k_1+\frac{14}{3}k_2+2kk_1k_2^2+\frac{4}{3}k_2^3$.

\section{Learning the master formula for surfaces}\label{learnmaster}
For surfaces there is another conceivable approach to machine learning cohomology formulae which is based on the master formula~\eqref{master}. We can attempt to set up a network which learns the irreducible, negative self-intersection divisors $C$ which determine the structure of this formula.

\subsection{Network structure}
\begin{figure}[h!!!]
\begin{center}
\includegraphics[width=15cm]{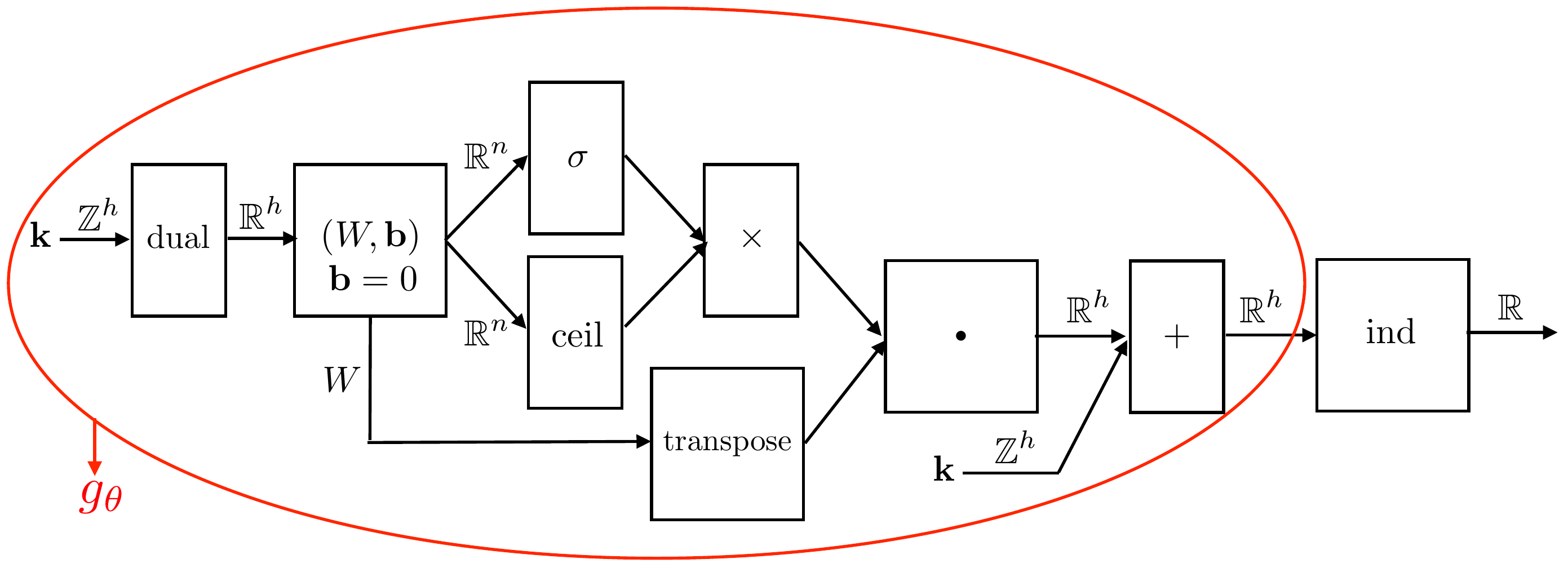}
\caption{\sf Structure of the network to learn divisors in master formula~\eqref{master}. The layers named dual and ind take the dual with respect to the intersection form according to Eqs.~\eqref{intdP} and compute the index from Eq.~\eqref{inddP}, respectively.}
\label{fig:mfl}
\end{center}
\vskip -5mm
\end{figure}

Fig.~\ref{fig:mfl} shows the structure of the network.
The part of the network denoted $g_\theta$, where $\theta=(W,{\bf b}=0)$, is a straightforward realisation of the master formula, that is, it converts an input divisor $D$ into a divisor $\tilde{D}$, as prescribed by Eq.~\eqref{master}. The only trainable element in this network is the linear layer with $n$ neurons. Its weights represent the divisors $C$ which appear in the master formula while the biases have been fixed to zero. (The network $g_\theta$ assumes that $C^2=-1$, in order to simplify Eq.~\eqref{master}. This assumption is indeed satisfied for del Pezzo surfaces to which we will apply the network.) Of course, we may not know, a priori, how many divisors are required so we will have to vary the width $n$ of the network and choose an optimal configuration. 

Recall from Eq.~\eqref{h0ind} that the zeroth cohomology dimension of the line bundle ${\cal O}_X(D)$ can be computed from the index of ${\cal O}_X(\tilde{D})$. For this reason we have added a further layer, called ``ind', at the end of the network in Fig.~\ref{fig:mfl} which simply computes the index. For del Pezzo surfaces this is done using the right-hand-side of Eq.~\eqref{inddP}.

The logistic sigmoid layer in Fig.~\ref{fig:mfl} consists of a logistic sigmoid function
\begin{equation} 
\sigma(x)=\frac{1}{1+e^{-x/\nu}}
\end{equation}
with adjustable width $\nu$. A value $\nu<1$ leads to a better approximation to the Heaviside theta function which appears in the master formula~\eqref{master}, than  $\nu=1$. On the other hand, $\nu$ should not be too small or else training becomes inefficient. It turns out that $\nu=1/4$ is a good compromise choice which we adopt in the following.

We can now train this network with data for the zeroth cohomology dimensions on del Pezzo surfaces, provided in the standard form~\eqref{traindata}. As mentioned, we will do this for various values of the width $n$ and choose a configuration with a large success rate. For such a trained network, we then read out the weights $W$ which, rounded to the nearest integer, should correspond to the divisors $C$ which enter the master formula.  

\subsection{Examples}
We will now train the above network for the cases of $\dps{r}$, where $r=1,2,3$. Line bundles ${\cal O}({\bf k})$ are labelled by $(r+1)$-dimensional integer vectors ${\bf k}=(k_0,k_1,\ldots ,k_r)$ and divisors are written as $D=k_0l+k_1e_1+\cdots +k_re_r$, where $l$ is the hyperplane class and $e_i$ are the classes of the exceptional divisors. 

For $\dps{1}$, $\dps{2}$ and $\dps{3}$, we use training boxes of size $k_{\rm max}=15,10,10$ and find optimal fits for network widths $n=1,3,6$, respectively. After reading out the weights and rounding to the nearest integer we find
\begin{equation}
\begin{array}{lll}
 \dps{1}:\quad&W\simeq \left(0,1\right)&\begin{array}{lll}D_1&=&e_1\end{array}\\[3mm]
 \dps{2}:&W\simeq\left(\begin{array}{rrr}0&1&0\\0&0&1\\1&-1&-1\end{array}\right)&\begin{array}{lll}D_1&=&e_1\\D_2&=&e_2\\D_3&=&l-e_1-e_2\end{array}\\[8mm]
 \dps{3}:&W\simeq\left(\begin{array}{rrrr}0&1&0&0\\0&0&1&0\\0&0&0&1\\1&-1&-1&0\\1&-1&0&-1\\1&0&-1&-1\end{array}\right)\quad&
            \begin{array}{lll}D_1&=&e_1\\D_2&=&e_2\\D_3&=&e_3\\D_4&=&l-e_1-e_2\\D_5&=&l-e_1-e_3\\D_6&=&l-e_2-e_3\end{array}
\end{array}
\end{equation} 
For each case, these are precisely all the self-intersection $-1$ divisors which enter the master formula~\eqref{master}. \\[2mm]
For higher del Pezzo surfaces $\dps{r}$, where $r>3$, the number of divisors $C$ which enter the master formulae increases further. For example, for $\dps{4}$ there are $10$ such divisors. For such cases, it is difficult to obtain all divisors $C$ from a single trained network.

A simple observation about the symmetry of the problem helps in dealing with those more complicated cases. Since the exceptional divisors $e_i$ are in generic positions, it is clear that the cohomology dimensions $h^0({\cal O}_{\dps{r}}({\bf k}))$ are unchanged under permutations of the corresponding integers $(k_1,\ldots ,k_r)$. Hence, our training sets have a redundancy which is described by the action of the symmetric group $S_r$. Suppose that we define the reduced training sets
\begin{equation}
 \left\{ {\bf k}\rightarrow h^0({\cal O}_{\dps{r}}({\bf k}))\,|\, |k_i|\leq k_{\rm max}\,,\; \;k_{i_1}\leq k_{i_2}\leq\cdots\leq k_{i_r}\right\} \eqlabel{dPts}
\end{equation} 
where $(i_1,i_2,\ldots ,i_r)$ is a permutation of $(1,2,\ldots ,r)$. There are $r!$ such sets, one for each permutation, in each of which the aforementioned redundancy has been removed. 

It turns out that training the network in Fig,~\ref{fig:mfl} with the non-redundant training sets~\eqref{dPts} is significantly more efficient than using their redundant counterpart. For 
$\dps{4}$, this observation saves the day. We can train $24=4!$ networks, one for each of the non-redundant training sets~\eqref{dPts}, and collect all (near) integer weights from these $24$ networks. Carrying this out for a training box with size $k_{\rm max}=7$ leads to success rates $\geq 0.97$ for all $24$ networks and provides all $10$ divisors $C$, which are given by
\begin{equation}
 e_i\;,\quad l-e_i-e_j\; ,
\end{equation} 
where $i,j=1,2,3,4$ and $i<j$.

\section{Conclusion}\label{conclusion}
In this paper, we have studied several approaches to machine learning of line bundle cohomology dimensions, both on complex surfaces and three-folds. 

Standard function learning of line bundle cohomology with fully-connected one or two hidden layer networks and logistic sigmoid activation functions can be implemented successfully. It can lead to large success rates (the percentage of line bundle cohomologies in the training box reproduced correctly after rounding to the nearest integer) of 90\% or sometimes higher.

However, in practice, the usefulness of such networks is limited. Training becomes more difficult for large cohomology dimensions, which is particularly relevant for three-fold examples. Even large success rates of, say, 90\% are not sufficient for reliable predictions when many line bundles are involved, as is typically the case in string model building. Finally, the trained network does not reveal any of the underlying mathematical structure of line bundle cohomology.\\[2mm]
Recent work~\cite{papermath,paperex} suggests and, in some cases, proves, that line bundle cohomology can be described by piecewise polynomial formulae, where the degree of the polynomials involved equals the complex dimension of the manifold. We have used this observation as a starting point for setting up a conjecture-generating network capable of learning such piecewise polynomial formulae. The structure of this network is shown in Fig.~\ref{fig:fl}.

The key observation is that it is sufficient to learn the various regions in the Picard lattice. Once these regions are known a simple fit determines the associated polynomials. We have carried this out successfully for surfaces, including for $\dps{1}$ and $\dps{2}$ where the cohomology formula was already known~\cite{paperex} as well as for some complete intersection surfaces whose cohomology formulae has not been previously worked out.

In the case of three-folds, the network managed to learn the known formulae for the zeroth and first cohomology of the bi-cubic CICY~\cite{Constantin:2018hvl} as well as a formula for the zeroth cohomology of another CICY manifold, which has only recently been worked out by different methods in Ref.~\cite{Larfors:2019sie}.\\[2mm]
As an alternative approach to conjecture-generating, we have designed a network (shown in Fig.~\ref{fig:mfl}) which learns the irreducible, negative self-intersection divisors $C$ which appear in the master formula~\eqref{master} for line bundle cohomology on surfaces. We have shown that this works straightforwardly for examples with a relatively low rank of the Picard group and a small number of weights, including for the cases $\dps{1}$, $\dps{2}$ and $\dps{3}$. For higher del Pezzo surfaces $\dps{r}$, where $r>3$, we can use a symmetry of the problem - the independence of the cohomology dimension on permutations of the exceptional divisors - to improve training efficiency. In this way, we have obtain all $10$ divisors $C$ for $\dps{4}$.\\[2mm]
The main purpose of this paper was to provide proof of concept that generating conjectures for line bundle cohomology formulae via machine learning is feasible. It would be interesting to implement these methods in a more standard fashion which allows generating cohomology formulae for large classes of manifolds. This knowledge might help getting to a better understanding of the mathematical structure underlying line bundle cohomology and assist the search for a more general version of the master formula.

\section*{Acknowledgments}
C.~R.~B. is supported by an STFC studentship and R.~D.~by a Skynner Fellowship of Balliol College, Oxford.



\begin{thebibliography}{99}
\ifx\doiref\asklfhas\newcommand{\doiref}[2]{\href{http://dx.doi.org/#1}{#2}}\fi
\raggedright 
\ifx\arxivref\asklfhas\newcommand{\arxivref}[2]{\href{http://arxiv.org/abs/#1}{arXiv:#1}}\fi
\raggedright

\bibitem{He:2017set}
  Y.~H.~He,
  ``Machine-learning the string landscape,''
 \textsf{\doiref{doi:10.1016/j.physletb.2017.10.024}{Phys.\ Lett.\ B {\bf 774} (2017) 564.}}
    
\bibitem{He:2017aed}
  Y.~H.~He,
  ``Deep-Learning the Landscape,''
 \textsf{\arxivref{1706.02714}}.
    
\bibitem{Ruehle:2017mzq}
  F.~Ruehle,
  ``Evolving neural networks with genetic algorithms to study the String Landscape,''
  \textsf{\doiref{doi:10.1007/JHEP08(2017)038}{JHEP {\bf 1708} (2017) 038}, \arxivref{1706.07024}}.
    
\bibitem{Bull:2018uow}
  K.~Bull, Y.~H.~He, V.~Jejjala and C.~Mishra,
  ``Machine Learning CICY Threefolds,''
 \textsf{\doiref{doi:10.1016/j.physletb.2018.08.008}{Phys.\ Lett.\ B {\bf 785} (2018) 65}, \arxivref{1806.03121}}.
  
\bibitem{Erbin:2018csv}
  H.~Erbin and S.~Krippendorf,
  ``GANs for generating EFT models,''
 \textsf{\arxivref{1809.02612}}.
  
\bibitem{He:2018jtw}
  Y.~H.~He,
  ``The Calabi-Yau Landscape: from Geometry, to Physics, to Machine-Learning,''
  \textsf{\arxivref{1812.02893}}.
   
\bibitem{Cole:2018emh}
  A.~Cole and G.~Shiu,
  ``Topological Data Analysis for the String Landscape,''
  \textsf{\doiref{doi:10.1007/JHEP03(2019)054}{JHEP {\bf 1903} (2019) 054}, \arxivref{1812.06960}}.
  
\bibitem{Halverson:2019tkf}
  J.~Halverson, B.~Nelson and F.~Ruehle,
  ``Branes with Brains: Exploring String Vacua with Deep Reinforcement Learning,''
  \textsf{\arxivref{1903.11616}}.
  
  \bibitem{Constantin:2018otr}
A.~Constantin, \emph{{Heterotic String Models on Smooth Calabi-Yau
  Threefolds}}.
\newblock PhD thesis, Oxford U., 2018.
\newblock \href{https://arxiv.org/abs/1808.09993}{{\ttfamily 1808.09993}}.

\bibitem{Buchbinder:2013dna}
E.~I. Buchbinder, A.~Constantin and A.~Lukas, \emph{{The Moduli Space of
  Heterotic Line Bundle Models: a Case Study for the Tetra-Quadric}},
  \href{http://dx.doi.org/10.1007/JHEP03(2014)025}{\emph{JHEP} {\bfseries 1403}
  (2014) 025}, [\href{https://arxiv.org/abs/1311.1941}{{\ttfamily 1311.1941}}].
  
\bibitem{Constantin:2018hvl}
  A.~Constantin and A.~Lukas,
  ``Formulae for Line Bundle Cohomology on Calabi-Yau Threefolds,''
\textsf{\arxivref{1808.09992}}.

\bibitem{Larfors:2019sie}
  M.~Larfors and R.~Schneider,
  ``Line bundle cohomologies on CICYs with Picard number two,''
 \textsf{\arxivref{1906.00392}}.
 

\bibitem{Klaewer:2018sfl}
  D.~Klaewer and L.~Schlechter,
  ``Machine Learning Line Bundle Cohomologies of Hypersurfaces in Toric Varieties,''
\textsf{\doiref{doi:10.1016/j.physletb.2019.01.002}{Phys.\ Lett.\ B {\bf 789} (2019) 438}, \arxivref{1809.02547}}.
  
\bibitem{papermath} 
C.~Brodie, A.~Constantin, R.~Deen, A.~Lukas, ``Topological Formulae for Line Bundle Cohomology on Surfaces'',  \textsf{\arxivref{1906.nnnnn, recently submitted}}.

\bibitem{paperex}
C.~Brodie, A.~Constantin, R.~Deen, A.~Lukas, ``Index Formulae for Line Bundle Cohomology on Complex Surfaces'',  \textsf{\arxivref{1906.nnnnn, recently submitted}}.

\bibitem{hartshorne} 
R.~Hartshorne, ``Algebraic Geometry", Springer Science \& Business Media, 2013.

\bibitem{gh}
P.~Griffiths, J.~Harris, ``Principles of Algebraic Geometry", John Wiley \& Sons, 2014.

\bibitem{huebsch}
T.~H\"ubsch, ``Calabi-Yau Manifolds: A Bestiary for Physicists", World Scientific, 1994.

\bibitem{Blumenhagen:2010pv}
  R.~Blumenhagen, B.~Jurke, T.~Rahn and H.~Roschy,
  ``Cohomology of Line Bundles: A Computational Algorithm,''
 \textsf{\doiref{10.1063/1.3501132}{J.\ Math.\ Phys.\  {\bf 51} (2010) 103525}, \arxivref{1003.5217}}.

\bibitem{Anderson:2007nc}
  L.~B.~Anderson, Y.~H.~He and A.~Lukas,
  ``Heterotic Compactification, An Algorithmic Approach,''
\textsf{\doiref{doi:10.1088/1126-6708/2007/07/049}{JHEP {\bf 0707} (2007) 049}, \arxivref{hep-th/0702210}}.

\bibitem{Gray:2007yq}
  J.~Gray, Y.~H.~He, A.~Ilderton and A.~Lukas,
  ``A New Method for Finding Vacua in String Phenomenology,''
\textsf{\doiref{doi:10.1088/1126-6708/2007/07/023}{JHEP {\bf 0707} (2007) 023}, \arxivref{hep-th/0703249}}.

\bibitem{Anderson:2008uw}
  L.~B.~Anderson, Y.~H.~He and A.~Lukas,
  ``Monad Bundles in Heterotic String Compactifications,''
\textsf{\doiref{doi:10.1088/1126-6708/2008/07/104}{JHEP {\bf 0807} (2008) 104}, \arxivref{0805.2875}}.

\bibitem{He:2009wi}
  Y.~H.~He, S.~J.~Lee and A.~Lukas,
  ``Heterotic Models from Vector Bundles on Toric Calabi-Yau Manifolds,''
\textsf{\doiref{doi:10.1007/JHEP05(2010)071}{JHEP {\bf 1005} (2010) 071}, \arxivref{0911.0865}}.

\bibitem{Anderson:2009mh}
  L.~B.~Anderson, J.~Gray, Y.~H.~He and A.~Lukas,
  ``Exploring Positive Monad Bundles And A New Heterotic Standard Model,''
\textsf{\doiref{doi:10.1007/JHEP02(2010)054}{JHEP {\bf 1002} (2010) 054}, \arxivref{0911.1569}}.

\bibitem{Blumenhagen:2008zz}
  R.~Blumenhagen, V.~Braun, T.~W.~Grimm and T.~Weigand,
  ``GUTs in Type IIB Orientifold Compactifications,''
 \textsf{\doiref{doi:10.1016/j.nuclphysb.2009.02.011}{Nucl.\ Phys.\ B {\bf 815} (2009) 1}, \arxivref{0811.2936}}.

\bibitem{Candelas:1987kf}
  P.~Candelas, A.~M.~Dale, C.~A.~Lutken and R.~Schimmrigk,
  ``Complete Intersection Calabi-Yau Manifolds,''
  \textsf{\doiref{10.1016/0550-3213(88)90352-5}{Nucl.\ Phys.\ B {\bf 298} (1988) 493}}.
  
\bibitem{Green:1986ck}
  P.~Green and T.~Hubsch,
  ``Calabi-Yau Manifolds As Complete Intersections In Products Of Complex Projective Spaces,''
  \textsf{\doiref{10.1007/BF01205673}{Commun.\ Math.\ Phys.\ {\bf 109} (1987) 99}}.

\bibitem{Kingma:2014vow}
  D.~P.~Kingma and J.~Ba,
  ``Adam: A Method for Stochastic Optimization,''
\textsf{\arxivref{1412.6980}}.

\bibitem{uap}
G.~Cybenko,
\textsf{\doiref{10.1007/BF02551274}{Math. Control Signal Systems (1989) 2: 303}}.

\end{thebibliography}
\end{document}